\documentclass{article}

\usepackage{arxiv}
\usepackage{url}
\usepackage{booktabs}
\usepackage{amsmath}
\usepackage{tabularx} 
\usepackage{longtable}
\usepackage{xcolor}
\usepackage{sidecap}
\usepackage{listings}
\usepackage{xspace}
\usepackage{graphicx} 

\definecolor{keyword}{rgb}{0.0, 0.0, 0.6}
\definecolor{comment}{rgb}{0.3, 0.6, 0.0}
\definecolor{string}{rgb}{0.58, 0.0, 0.82}

\newcommand{\sys}{\textbf{DataInquirer}\xspace }
\begin{document}

\title{A System for Quantifying Data Science Workflows with Fine-Grained Procedural Logging and a Pilot Study}

\author{
Jinjin Zhao \\
University of Chicago\\
j2zhao@uchicago.edu\\
\And
Avigdor Gal\\
Technion -- Israel Institute of Technology\\
avigal@technion.ac.il
\And
Sanjay Krishnan\\
University of Chicago\\
skr@uchicago.edu
}



\maketitle

\begin{abstract}
It is important for researchers to understand precisely how data scientists turn raw data into insights, including typical programming patterns, workflow, and methodology.
This paper contributes a novel system, called \sys, that tracks incremental code executions in Jupyter notebooks (a type of computational notebook).
The system allows us to quantitatively measure timing, workflow, and operation frequency in data science tasks without resorting to human annotation or interview.
In a series of pilot studies, we collect 97 traces, logging data scientist activities across four studies.
While this paper presents a general system and data analysis approach, we focus on a foundational sub-question in our pilot studies: How consistent are different data scientists in analyzing the same data? 
We taxonomize variation between data scientists on the same dataset according to three categories: semantic, syntactic, and methodological. 
Our results suggest that there are statistically significant differences in the conclusions reached by different data scientists on the same task and present quantitative evidence for this phenomenon.
Furthermore, our results suggest that AI-powered code tools subtly influence these results, allowing student participants to generate workflows that more resemble expert data practitioners.
\end{abstract}

\section{Introduction}
Data science is a rapidly growing discipline affecting academia, science, and industry.
Data scientists are using ever more sophisticated software tools for data analysis.
While some view the quantitative methodology as an objective scientific process, human judgment influences every stage of the data science project lifecycle.
For many decades, researchers have argued that it is important to understand the human element of data science, namely, \emph{how people turn raw data into insights}~\cite{tukey1977exploratory, battle2023exactly}.
Researchers have evolved this question to understand typical programming patterns, workflow, and methodology in the data science era.
New academic communities have arisen around this research area to identify how to optimize tools for human factors~\cite{hilda}.

Today, this research is primarily driven by user interview studies of data scientists or coarse metrics about tool use.
For example, a seminal study by Kandel et al. still shapes how the community thinks about how data are preprocessed, visualized, and explored data~\cite{kandel2012enterprise}.
The challenges highlighted by Kandel et al. around data quality and profiling are still referenced a decade later in studies of the current state of data science~\cite{paleyes2022challenges}.
However, both the HCI literature~\cite{kjeldskov2003review, kujala2000effective, kosara2003thoughts, wright2005user} and the research methods literature~\cite{weiss1995learning} acknowledge the difficulties and time-investment needed to collect high-quality information from interview studies of users.
Thus, in a rapidly changing software ecosystem, interview studies conducted on relatively small samples of willing experts can drive the community’s model for user behavior for years.

As open-access to research artifacts has become more prevalent, e.g., through Github, there have been a handful of recent studies that seek to quantitatively measure data science practice~\cite{kallen2020jupyter, ramasamy2023workflow, raghunandan2023code, wang2020better, pimentel2019large}.
These approaches leverage code analysis over existing public repositories of data science code.
A post-hoc analysis of such code can be quite challenging.
For example, a 2019 study by Rehman~\cite{rehman2019towards} had to filter a corpus of 1.25 million code artifacts to about 50,000 that could be reproduced and analyzed. 
Such results point to a larger issue of measuring data science in a post-hoc way: researchers are at the mercy of the software engineering practices of the data scientists who created the code.
As researchers from the database systems community, we offer a new quantitative tool for such researchers: a way to proactively log real data science in progress.

Computational notebooks have emerged as a dominant interface in data science~\cite{kluyver2016jupyter}.
Computational notebooks are collections of executable {\em cells} that intersperse code, text, and results. 
Each notebook has a {\em kernel} that hosts an interactive code interpreter, which state persists across a user's session.
Cells can be selectively executed by passing code cells to this kernel and returning the result to the notebook.
There is no particular order in which cells have to be executed.
Every major cloud provider has a notebook interface, such as Google's Collaboratory and AWS's Athena~\cite{colab, athena}.
A missing piece from this software stack \emph{fine-grained telemetry}, where data scientist actions are logged for further analysis.
This log can be analyzed to identify key events and design decisions made by a data scientist.
Such data is invaluable for understanding the final data products produced by data scientists and their intents and workflows.

This paper contributes a novel system, called \sys, that tracks incremental code executions in Jupyter notebooks (an open-source computational notebook).
The system is based on an execution log that captures every cell execution and its result throughout user sessions.
This log can be used to reconstruct the timeline of operations and the kernel's internal state over a data science project.
The log is durable over multiple kernels.
The system allows us to quantitatively measure timing, workflow, and operation frequency in data science tasks without using human annotation, interview studies, or analysis of a version-control system like Github.
The system further shows how a query interface over these logs can support different notebook analysis questions.

\subsection{Pilot Study}
To demonstrate a proof-of-value, we use \sys to measure the \emph{variation across data scientists on the same task}.
This question is of deep interest to the data science community.
For example, would two different data scientists analyzing the same data arrive at different conclusions?
Variation is not necessarily negative; the research literature establishes that individual variation and diversity are crucial in collective deliberative processes~\cite{bohman2006deliberative}.
For example, a group of data scientists may collectively discover different data anomalies using different data cleaning techniques.

In our study, we provide a set of data scientist subjects with the same dataset, analysis instructions, and coding environment.
We use the logs collected from \sys to compare and contrast how different the data scientists approach the same problem.
A fine-grained execution log can give insights into skill differences (e.g., due to timing) and process differences (e.g., error resolution approaches) that may not be obvious from comparing final products.
Quantifying data scientist variation is the first step towards building robust decision-making systems that can account for differences in process or result due to data scientist judgment.
We found that beyond using different coding styles, different data scientists can make data cleaning or standardization choices that are semantically incompatible.
\textbf{Over all data collected a consistent theme emerged, where data scientists vary more in how they scope a data science problem than in how they complete a data science problem}.

Serendipitously, during our user studies, AI-driven coding tools dramatically increased in popularity.
This led to a natural experiment of determining how data scientist behavior changes when offered the assistance of AI-powered coding tools. 
Our fine-grained logs paint an interesting picture where AI-based tools reduce the coding burden in a data science task.
These results beg further investigation on whether AI-based coding tools reduce the potential positive effects of variation (e.g., diversity of conclusions).

\subsection{Organization and Main Findings}
The goal of this paper is two-fold: first, to present a logging system for computational notebooks, and second, to present results from user studies to evaluate differences in data scientist behavior on the same task.
This pilot task illustrates some research questions that can be answered with such fine-grained telemetry. This paper is structured as follows:
\begin{itemize}
    \item We introduce \sys and describe the scope and capabilities of the execution logging system. We also outline \sys's query systems and its ability to replay intermediate states of a data workflow (Section 2).
    \item We motivate the pilot study of quantifying variation between data scientists. We introduce datasets on data scientist workflows collected over three years of research (Section 3).
    \item We formalize our definitions of variance and discuss the methodology for analysis (Section 4). 
    \item We present findings from our dataset on various research questions, emphasizing the variance between participants (Section 5). 
    \item We position \sys in the broader research agenda of understanding data science (Section 6).
\end{itemize}

To summarize the main findings of the paper,
\begin{itemize}
    \item When given the same data science task on dirty data, participants identify different error types and use different methods to fix those errors.
    \item Using different data-cleaning methodologies is associated with divergent conclusions on at least one of the pilot study datasets.
    \item We find that experts are more concise and faster in their approach to a data science task. We find that the gap between novices and experts significantly narrows after the proliferation of LLM-based coding tools in 2023.
    \item Based on participant self-reporting on LLM use, we observe a ``funneling'' effect where LLMs may guide participants down certain analysis workflows.
\end{itemize}
\section{\sys: Architecture and Description}
\begin{figure}[ht!]
    \centering
    \includegraphics[width=0.6\linewidth]{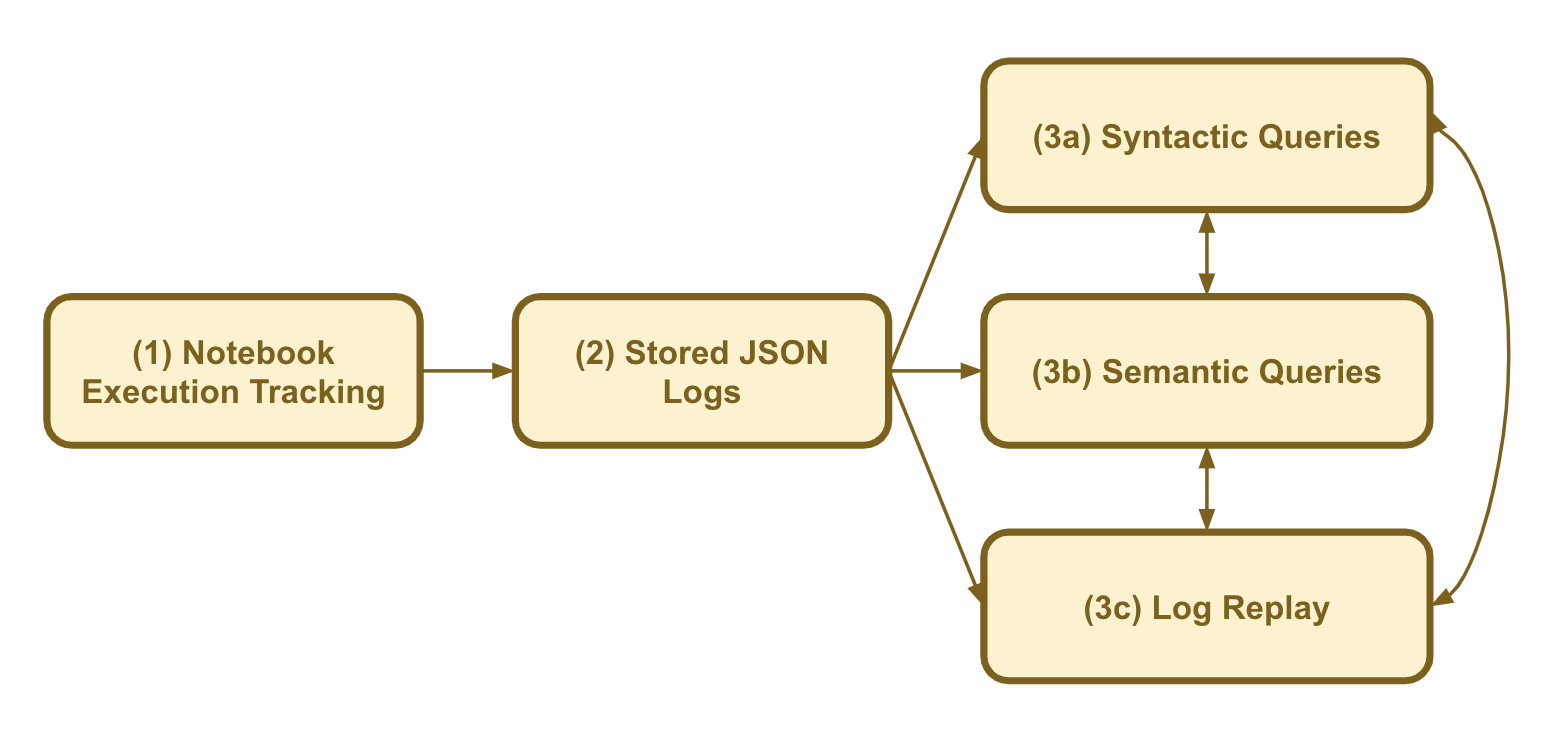}
    \caption{Outline of the components of \sys}
    \label{fig:sys}
\end{figure}

\sys is an end-to-end system for logging and analysis of computational notebooks. The architecture of \sys is shown in Figure \ref{fig:sys}. The system comprises three parts - a logging framework that captures code and additional statistics at runtime, a JSON-format document store, and multiple auxiliary libraries for more complex analysis. The auxiliary libraries include components for string-based syntactical queries, abstract syntax tree-based semantic queries, and execution replay of the log history. While many of the techniques used within this system bear similarities to previous work, to the best of our knowledge, this is the first system that incorporates all such techniques to provide a comprehensive solution for notebook code tracking and analysis.
We focus the paper's discussion on Python notebooks as they are the most popular \cite{microsoft_2019}, but techniques are fully compatible with other interpreted languages such as R or Julia.

\subsection{Anatomy of a Computational Notebook}
Jupyter Notebook is an example of a computational notebook, an open-source web application that allows one to create and share documents that contain live code, equations, visualizations, and narrative text. It is widely used in data science, research, and education for interactive computing and data analysis. One of the key features of the Jupyter notebook is its support for interactive computing. We can execute code in individual cells and see the results immediately below the cell. This allows for an iterative and exploratory workflow where one can quickly prototype code, visualize data, and analyze results.

Jupyter Notebook follows a client-server architecture. When opening a Jupyter Notebook in a web browser, interaction is performed with the client interface. The server manages and runs the code and communicates with the kernels. The code entered into the browser interface triggers a message to the server. The kernel is a computational engine that executes the code contained in the notebook document. Jupyter supports multiple kernels, including IPython (for Python), Julia, and R. Each kernel is an isolated program. The server passes messages to the kernel to execute. The notebook maintains a stateful session with the kernel, meaning that variables and other objects created in one cell are accessible in subsequent cells, supporting the building of complex analyses step by step.

Jupyter uses a JSON-based messaging protocol for communication between the client interface and the kernel. When a code cell is executed in the notebook, the client sends the code to the kernel over a network connection. The kernel then executes the code and returns the result to the client, which displays it in the notebook interface.

\subsection{Execution Logging}
The first module of \sys is an execution log. Note that \sys is a logical logging scheme that captures the code executed and not the state of the kernel.
Internal variables, such as the data the data scientist is working with or models derived from that data, are not logged. 
Such data are reconstructed by replaying the logs.
A first requirement is that the kernel is executed in an isolated container where filesystem accesses, system calls, and the Python environment can be controlled.
We first record the execution environment, which includes the external library dependencies and metadata about the notebook file.

Logs are collected throughout a user's data science session using the following protocol: logging is triggered when the user prompts the notebook interface to execute a code cell. This code can be a new cell, an existing cell, or a modified existing cell.

\begin{itemize}
    \item Execution-Ahead-Log. \sys captures the message between the client and the server. Prior to execution, it logs the intended code to run.
    \item Execution-Log. The server passes the code to the kernel to execute, and the kernel returns a result (or an exception). \sys captures stdout and stderr results as well.
\end{itemize}

With this logging protocol, the entire state of the execution is captured. Even if the kernel fails or runs into an exception, the code that triggered the exception is logged. Logs are synchronously stored on disk, so there is a durable ledger of executions throughout the user session. In our first version, we assume single-user computational notebooks. That is, a single data scientist is associated with each notebook.

\subsubsection{Logging Format}
Some careful choices were made about the format of stored data. These choices greatly accelerated the timeline of analysis performed over the logs. Firstly, absolute IDs were defined over the logs at various granularities. These IDs persist through all operations using \sys's analysis modules so that any generated data items can reference the original log data. The IDs are as follows:

\begin{itemize}
\item \textbf{Document ID.} Each participant is associated with a unique document identifier. Metadata related to the entire workflow, such as the number of logs and summary statistics, are associated with this document. 
\item \textbf{Log ID.} Each execution of a Jupyter cell is associated with a log ID. The ID for each cell is assigned incrementally based on the order in which the cells are processed. Metadata related to cells, such as execution time and error statements, are associated with this ID.
\item \textbf{Line ID. } Finally, individual lines of code within a cell are also given an ID. This is the smallest granularity by which we organize the data. There are multiple intuitive ways to organize code; for example, the abstract syntax tree (AST) model represents code as a graph. Instead of dealing with heterogeneous data representation, we translated the metadata generated from this representation into line granularity. For metadata that span multiple lines, we duplicated the information across different line IDs. Metadata that span a partial line are linked to the entire line - we found that this approximation works in practice. If an approximation is not appropriate, additional positioning metadata can be included. Some examples of metadata include extracted comments, extracted comment-free code, and code categorization tags.

\end{itemize}

\begin{figure}[ht!]
    \centering
    \includegraphics[width=0.5\linewidth]{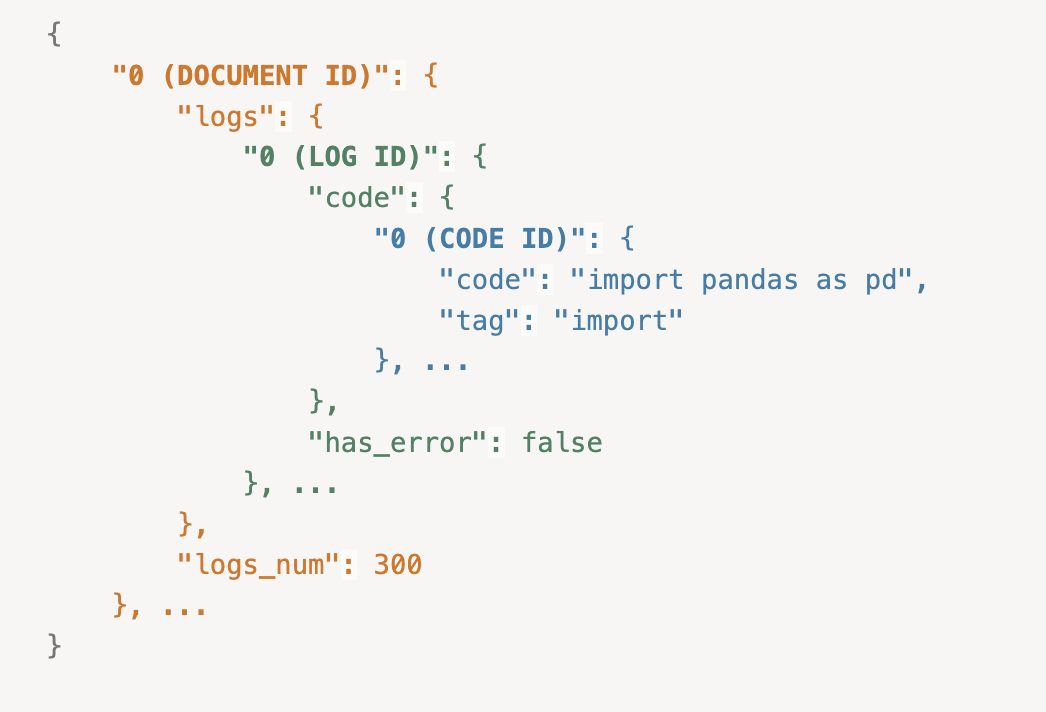}
    \caption{Example of the stored log structure}
    \label{fig:log}
\end{figure}

The log data is stored in a JSON document structure. Figure \ref{fig:log} shows an example of this structure. Participant documents are first-level objects; logs are contained within participant documents, and code lines are contained within logs. The natural hierarchy of the three different data objects is persevered. Our system easily adds key-value metadata to the log structure by linking it to specific IDs as analysis is performed. This allows cumulative analysis where we can use previously processed metadata for further analysis. Previous logging structures have focused on a linear log history for notebooks but have not explored integration with metadata storage \cite{Head2019ManagingMI, mack2020lineage, kluyver2016jupyter}.

\subsection{Queries Over The Logs}
In \sys, the logs are fully queryable. These queries allow us to analyze the code and process the data scientist uses. \sys supports two types of queries: literal queries and syntactic queries. Queries can be further composed and nested.

When applicable, the query output returns the same log structure as the original data output. The data returned is the subset of original data that matches the conditions of the query request \cite{views}. Therefore, further queries or additional analysis can be applied to results using the same libraries as the analysis over the original log.

\subsubsection{Literal Query}
A literal query is simply a full-text search over items, i.e., identifying a literal code pattern. This has similarities to standard document queries with added keywords \cite{mongo}. These queries over the logs provide the ability to perform many complicated operations. An example of this query in Python syntax is shown: 

\begin{lstlisting}
    find(["AND", (participant_key.ID, 1), 
            (line_key.code, ["CONTAINS", "np"])])
\end{lstlisting}

 Our example query returns the lines of code from the participant with \texttt{id = 1} that contains the substring \texttt{np}. \sys supports regular expression in addition to substrings and nested combinations of logical operators ("AND", "OR", "NOT). We explicitly also support string search with the "CONTAINS" operators that search for the existence of a substring within properties. We implement custom keywords ``\texttt{participant\_key}'', ``\texttt{log\_key}" and ``\texttt{line\_key}''. When a query is made with the ``
\texttt{participant\_key.property}'' key, our API matches all participant documents that satisfy the condition on that property; \texttt{log\_key.property} and \texttt{line\_key.property} keys function similarly. In other words, these keywords are both a shorthand for nested properties and allow us to treat the list of absolute keys that are sub-properties of those properties as an array.

Now, let us discuss the use of string search for our variation analysis. A direct string search does not account for all possibilities for where a substring may be used. For example, '\texttt{plt}' is a standard alias that references the ``\texttt{matplotlib.pyplot}'' library. We expect that searching for this substring shows when functions in that library are called. However, a participant is technically allowed to use any alias for the library, so that keyword may not appear in their code. Additionally, the '\texttt{plt}' substring can be a part of another string or be set to be a different variable. However, these string searches give us a good approximation over aggregate data, and we found that adjusting the variations in the string search allowed us to minimize the edge cases.

\subsubsection{Syntactic Queries}
It is easily seen that literal queries may not be sufficient for all types of code analysis. We might be interested in what the code structure and relationships between elements can inform us. This is where syntactic queries come in, as they interpret the code in the execution log as an abstract syntax tree (AST). 

An AST represents the code as a tree structure, capturing various logical aspects of the program. Each node in the tree corresponds to a logical construct, such as a variable assignment, a function definition, or an arithmetic operation. The tree's hierarchical nature reflects the nested and sequential structure of the source code. This hierarchical representation allows for a detailed understanding of the code's structure and behavior that we cannot achieve with a simple string search.

\sys includes a module that uses AST to capture specific aspects of the code useful in analysis. Specifically, AST objects, top-level Python logic, and code lineage can be extracted, and custom code can be inserted into the AST graph. We document the first three cases here and discuss the last functionality regarding code replay in the following section.

There are various benefits to the functionalities of this model. Firstly, by directly querying for AST objects, we avoid any issues related to syntax, as discussed in the previous section. This allows us to capture more accurate queries. Indeed, we feel confident about the previous literal queries because we can compare them to equivalent, more complex AST-based results. For example, we can mitigate the previous case where `\texttt{plt}' can reference multiple meanings with AST modules. We can specify whether the term is used as a library alias by evaluating whether the `\texttt{plt}' string is contained in a \texttt{Import} AST object.

\sys can filter the logs for basic AST objects, as well as filter a line of code to return the outermost AST object. For example, the system can efficiently identify all library imports in a log or determine if a code line belongs within a function definition. This capability, combined with the ability to search for specific text using literal queries, enables comprehensive and nuanced log data analysis. The use of AST to capture code lineage is well-documented in past works \cite{grafberger2021mlinspect, microsoft_2019, mack2020lineage}. Code lineage identifies the execution and data dependencies between a code element and the preceding code elements. To obtain this lineage, we implement standard AST techniques. We annotate the base graph with edges between variable assignments and variable calls using a depth-first traversal of the original AST graph.

\sys enables querying logs for both forward and backward lineage. Given a specific code line index, we can return its forward and backward lineage. The forward lineage includes all preceding lines on which the specified line depends, while the backward lineage includes all subsequent lines on which it depends. Users can specify whether the returned lineage should include the top-level code logic. For instance, users can choose whether to return the entire loop if a parent line is within a for loop.
Additionally, a query can be modified to return only the direct parent or child dependencies. There are multiple reasons why understanding code lineage could be useful. Some use cases include capturing the dependency of imported libraries, understanding branching in the user's data flow, and tracking downstream data usage. Additionally, it aids in debugging, refactoring, and ensuring that changes in one part of the code do not inadvertently impact other parts. Understanding code lineage is essential for effective code management and ensuring robust software development practices \cite{microsoft_2019}.

\subsection{Log Replay}

The ability to replay a data scientist's actions is important for researchers trying to understand their process. Replay allows for the reproduction of the intermediate states of the workflow. Log replay aims to fast-forward and rewind through the log history and reconstruct the kernel state. We would like to do this efficiently and minimize the time between the request for replay and the intermediate state.

At initialization of replay, \sys creates a new iPython kernel with the same Python dependencies as the execution log. This kernel will run code in the order given by the logs. The naive method to replay to an intermediate state is to run the logs until the requested intermediate state is captured. However, replaying logs that occur later in history can be expensive because it requires rerunning all preceding logs. There has been some prior work in notebook live migration~\cite{li2023elasticnotebook}, but to the best of our knowledge, there has not been an exploration of arbitrary log replay to any historical state for notebooks.

As a practical matter, \sys implements a simple checkpointing optimization to optimize replay. At regular log intervals, the state of the kernel is materialized. During re-execution, the system finds and loads the closest checkpoint. As an optimization, \sys does not checkpoint the entire kernel state and simply checkpoints user-defined variables.
However, the stored variables do not include library imports and function declarations. Therefore, the system uses the query and AST modules to search all prior logs for import statements and function definitions. These code slices are also loaded into and executed for the replay kernel. The log query history is then re-executed until it reaches the requested log state.

The checkpointing procedure is efficiently implemented with a code re-writing module, which transforms the code that the user is executing. 
We implement a save function that uses the \texttt{pickle} format to save current active variables to file. 
The save function queries the current Python interpreter state to get all the active variables and uses AST analysis to determine the user-modified active variables. 
The latter are prioritized for checkpointing.
A call to this function is added to the original code during execution. 
All data objects in the local environment are saved to disk using our checkpoint function.  
 This is implemented similarly to the method discussed above. This method allows us to analyze data states, even in user patterns where the work is primarily nested.

The replay functionality was evaluated for accuracy over the expert logging dataset by matching the error statements from the logs with those generated during the replay. When there was no match, it was solely due to a difference in random seed or environment-specific variable naming. We found these errors inconsequential to our analysis, and all logs without original errors could be replayed with our dataset. 

\subsubsection{Discussion of Randomness}
Since \sys is a logical logging scheme, it captures the code executed and not the state of the kernel.
In principle, each log message has to be re-executed in sequence to reconstruct the notebook state at any time.
However, non-determinism in the executed code can lead to some issues:

\begin{enumerate}
    \item Programmatic Randomness: The output of random manipulations cannot be exactly replicated. For example, randomness may be used to construct the training-test split \cite{pimentel2019reproducibility}.
    \item Parallel computation: Replication of parallel programs is difficult since the order of operations can be variable. Many data science operations (such as machine learning modeling) leverage threaded execution interfaces that may not be deterministic \cite{bugnet2005}.
    
\end{enumerate}

In our first version of \sys, we ignore the impact of randomness. For our goal of analyzing processes in data science, the process should not depend on traits of data affected by random operations. Therefore, even if we do not recover the exact state due to variations from randomness, the state we achieve should be sufficient for executing the relevant cell. For example, suppose a data array is split with the \texttt{train\_test\_split} from the \texttt{scikit-learn} library. In that case, the cell re-execution and subsequent analysis should ideally be independent of the exact samples in the training and testing dataset.

\section{Pilot Study: In-Task Variation}
Next, we describe our pilot study based on \sys. 
This study aims to capture and examine detailed metrics about data analysis processes and to use these insights to inform larger studies to improve the development of data analysis tools and workflows.

\subsection{Motivation}
We motivate the topics addressed in this study. There is a lack of methodological consensus on defining many crucial steps (or their best practices) in data science.
In a Nature article about the reproducibility crises, MacCoun and Perlmutter raised an intriguing question about data analysis, namely, whether the judgments that humans make in tuning code, data collection, or cleaning could affect the statistical validity of their analyses~\cite{maccoun2015blind}.
Since there is no one right way to clean or correct data, they argued that proper science necessitated a ``blind'' analysis, where these judgments would have to be made independent of treatment/control assignments.
However, beyond confirmation bias and prejudicial data modification, data science is a diverse field drawing its heritage from many different disciplines~\cite{wing2006computational}. 
It is not inconceivable that different data scientists, with different training, perspectives, and prejudices, would approach the same task differently.
Without clear best practices, a basic question arises: how consistent are different data scientists with the same task? 
The answer to this question has wide-ranging implications from statistical validity (i.e., how do we build trust in a set of conclusions) to data governance (i.e., how do we structure data-science projects to be accurate, ethical, and reproducible).

The general topic of differences in opinion has been studied in psychology and decision sciences.
For example, Sunstein has long researched the statistics of group deliberations in terms of diversity, polarization, and consensus~\cite{sunstein2005group}
Black et al. further studied quantitative tools to measure group deliberations in a political context~\cite{black2014methods}.
This work is highly relevant, even if in a non-computational context.
Data science is fundamentally a means to an end for organizational decision-making.
Our pilot study provides a new data point of how different data-driven analyses may be.

Understanding this variation is also important for creating robust decision processes.
In his book ``Noise'', psychology Nobel laureate Daniel Kahneman describes how even a single decision maker can be inconsistent in the decisions that they make~\cite{kahneman2021noise}.
Kahneman points to decision variations that can have meaningful impacts on the stakeholders involved, e.g., a inconsistent judge between cases.
We argue that the same concern can happen in data science.
Consider a data series with a handful of missing measurements.
One data scientist might fill in those measurements with an imputation technique.
Another might drop those measurements altogether.
Both these options are reasonable with no other context. However, they can result in semantically different final datasets.
This raises an intriguing challenge to any data science project -- would choosing a different data scientist have resulted in a different conclusion?

\subsection{Collected Datasets}
Over three years, we captured collected logs of participants in four capturing tasks. The datasets described in this section have been anonymized, and are open and available for access. Through each iteration of this dataset collection, we have learned practical lessons about how to conduct logging in a data science education environment. These lessons are covered in the final discussion of the paper.

\vspace{0.5em}

\noindent \textbf{Student Participants.} The primary study participants consisted of senior undergraduate students majoring in computer science, each possessing over two years of experience with Python programming. All participants received uniform instructions for the assigned task, ensuring a standardized approach across the cohort. The task was integrated into their regular coursework, providing a familiar context for completion. Importantly, as stipulated by the Institutional Review Board (IRB), the research component of the task was designed to be entirely independent of their academic grading, thereby preventing any influence on their academic evaluation and ensuring voluntary and unbiased participation in the study.

\noindent \textbf{Expert Participants.} The expert participants in this study were professionals holding research positions in data science, either in academia or industry. Each participant self-identified as an "expert" in Python for data science, underscoring their advanced proficiency and extensive experience in the field. Consistent and detailed instructions were provided to all participants to ensure a standardized approach to the task. Recruitment was conducted through the author's professional networks, ensuring that the experts were not previously connected to the study or its outcomes, thereby minimizing potential biases and ensuring the integrity of the research findings.

\subsubsection{City of Chicago Traffic Analysis Dataset: Students only}
In our first experiment in 2021, the students were asked to work with the City of Chicago Traffic Analysis Dataset~\footnote{\url{https://data.cityofchicago.org/Transportation/Traffic-Crashes-Crashes/85ca-t3if}}. 32 students in the class were given the following prompt:
\begin{quote}
In this assignment, you will be exploring the City of Chicago traffic crash dataset. We want to learn more about the trends of crashes and visualize if there are any correlations between crashes and other factors.
\end{quote}
Students were prompted through a series of data exploration tasks, each of which is listed in Table \ref{tab:chicago}.
\begin{table}[ht!]
\centering
\begin{tabular}{|p{0.8\linewidth}|}
\hline
i. Many of the columns of this dataset contain "missing values." For each of the columns, describe how missing values are represented (the definition is up to you!) and how many missing values there are. \\ \hline
ii. Semantic errors are data errors where the data is present and syntactically correct but is questionable in context, e.g., a manager who earns significantly less than their direct subordinates in an employee database. Identify a column with semantic errors in this crash dataset. Consider values that don't make any sense (it's up to you to define this!), and read through the descriptions of the columns on the website. \\ \hline
iii. Erroneous or missing data often cannot be completely recovered, and sensible defaults must be used. For all of the errors that you found (missing and semantic errors), you must either remove that row from the dataset or replace the broken value with a sensible default. Describe your process. \\ \hline
Iv. Produce a chart that best conveys the relationship between the three condition columns and the frequency of crashes in this dataset. \\ \hline
v. Which neighborhoods in the city seem to have a higher frequency of crashes? \\ \hline
vi. What other variables might more directly contribute to crash frequency than neighborhood? How can you determine this? \\ \hline
\end{tabular}
\caption{Assignment questions given in the City of Chicago Traffic Analysis dataset}
\label{tab:chicago}
\end{table}
Using \sys, we tracked the code written by the students as well as self-reported answers to the above questions.

\subsubsection{Flight Delay Dataset v1: Students only} In 2022, the assignment was revised so that the students were given a machine-learning classification problem to generate a prediction model. 
Given the full 2018 Air Flight dataset ~\footnote{\url{https://www.kaggle.com/datasets/robikscube/flight-delay-dataset-20182022}}, participants were tasked with predicting flight departure delays exceeding 15 minutes on the 2019 version. Columns with information at or after departure were withheld. The dataset exhibited significant bias, with only 18.7\% of 2018 flights delayed. This made the machine learning problem particularly challenging: using the presented columns with minimal cleaning and standard scikit-learn models yielded predictions no better than a baseline of always predicting no delays \cite{zhao_2023_hilda}.
Unlike the previous study, participants were given a more open-ended task without step-by-step guidance.
We gathered full logs from 28 participants. Initial results on this particular dataset were discussed in our original vision paper.

\subsubsection{Flight Delay Dataset v2: Students, Experts}
In November 2022, ChatGPT was released \cite{openai2023chatgpt}. This changed the landscape of machine learning by making the newest advances in large language models (LLMs) accessible to the public \cite{farina2023adoption}. One area that saw large discussion was the use of these models by students for programming assignments \cite{Macneil2022TheIO}. We suspected that adopting LLMs would lead to quantifiable variation between students in 2022 and 2023 on the same data science task. Therefore, in 2023, we decided to track the same task over the Flight dataset in the same course, with the difference that students were allowed to use large language models to complete their assignment. We collected 33 full logs, as well as user responses on large language use.

Since large language model use is outside the Jupyter notebook environment, we explicitly asked students to respond to how they used these models. Table \ref{tab:llm} showed three aspects requested from the students - whether they used generative AI (LLM) tools, what prompts they requested, and what code they directly generated from these tools.

\begin{table}[ht!]
\centering
\begin{tabular}{|p{0.8\linewidth}|}
\hline
i. Please Respond with Yes or No: Did you use generative AI tools for this assignment? \\ \hline
ii. Please keep track of all the prompts you used in this section. Please include all parts of multi-prompt questions, as well as any prompts whose outputs you did not end up using. You should add the prompt you used to the list below. \\ \hline
iii. In a comment before any code, please also include a note if you used generative AI for the following lines of code. Specifically, please add a comment with the word "GENERATED" all in caps. If relevant, include the number for the relative corresponding prompt from the previous section in brackets. \\ \hline
\end{tabular}
\caption{Survey questions related to LLMs given in the Flight Delay Dataset v2 dataset}
\label{tab:llm}
\end{table}

\subsubsection{Flight Delay Dataset v3: Experts} Since the participants in the previous datasets were largely novices with little experience in data science, variations in their approach may differ from data scientists with more experience in the field. Gathering data from experienced data scientists can provide benchmarks for best practices, efficient workflows, and real-world problem-solving strategies. The same task given to students was also given to 4 industry experts. There was one slight modification to simplify the process; the experts did not need to apply their model to the 2019 dataset with unseen labels. Additionally, to minimize changes between this and the v1 dataset, they were asked not to use any LLM tools.

\section{Data Analysis Methods}

\subsection{Formal Characterization of Variation}
Before we discuss the analysis of the datasets, let us characterize what we mean by variation. In particular, we identify three key types of variation in data science workflows: syntactic, semantic, and methodological.

\subsubsection{Syntactic Variation}
Data scientists may employ varying programming patterns and use different libraries in their analysis code. \emph{Syntactic variation between two code segments $a$ and $b$ is when both $a$ and $b$ can produce the same result but are expressed differently in code.} Let us consider a simple example where we have a DataFrame `df`, and we want to filter the rows based on a condition and then select specific columns. 
\begin{lstlisting}
import pandas as pd

# Creating a sample DataFrame
data = {
    'Name': ['Alice', 'Bob', 'Charlie', 'David', 'Eve'],
    'Age': [24, 27, 22, 32, 29],
    'City': ['New York', 'Los Angeles', 'Chicago', 'Houston', 'Phoenix'],
    'Score': [85, 90, 88, 76, 95]
}

df = pd.DataFrame(data)
\end{lstlisting}
We aim to filter the DataFrame to include only rows where the `Age' is greater than 25 and then select the `Name' and `Score' columns.
We will perform this using two methods that ultimately produce the same result.
\begin{lstlisting}
# Using .loc to filter and select columns
result1 = df.loc[df['Age'] > 25, ['Name', 'Score']]
# Using .query to filter and then selecting columns
result2 = df.query('Age > 25')[['Name', 'Score']]
\end{lstlisting}

Both methods produce the same result: a data frame containing only the `Name' and `Score' columns for rows where the `Age' is greater than 25. 
Syntactic variation characterizes a similar concept of a paraphrase or a synonym in natural language. 

Understanding syntactic variation is crucial to organizing the code and data in notebooks.
If data scientists perform the same tasks with very different syntactic patterns, it complicates code searchability since the same operations might be represented with very different literal programs. 

\subsubsection{Semantic Variation}
Semantic characterizes how data scientists may arrive at semantically different data products given the same instructions. Let $a$ and $b$ be two code segments that conform to the same specification $s$. \emph{Semantic variation between the two code segments $a$ and $b$ is when both $a$ and $b$ produce different results.} 

Consider an example with data imputation. We are given a dataset that looks as follows:
\begin{lstlisting}
import pandas as pd
import numpy as np

# Creating a sample DataFrame with NA values
data = {
    'Name': ['Alice', 'Bob', 'Charlie', 'David', 'Eve'],
    'Age': [24, np.nan, 22, 32, np.nan]
}

df = pd.DataFrame(data)
\end{lstlisting}
The task is to calculate the median age. Given the `NA' values, this task is under-specified, and the data scientist must pick an approach to solve it.
One approach is to fill the `NA' values using the `.fillna()'' method. For simplicity, we will fill numerical columns with the mean of the column and categorical columns with a placeholder string. Another is to drop any rows that contain `NA' values using the `.dropna()' method.
\begin{lstlisting}
# Filling NA values
filled_df = df.copy()
filled_df['Age'].fillna(filled_df['Age'].mean(), inplace=True)
# Dropping rows with NA values
dropped_df = df.dropna()
\end{lstlisting}
It is easy to confirm that the first approach returns a result of $26$, and the second result returns a result of $24$. 
Understanding semantic variation is important as it is ultimately a form of noise.
If data scientists using the same data use incompatible assumptions, the final results may be statistically invalid.

\subsubsection{Methodological Variation}
Data scientists vary in the time-ordering or time-taken of key steps in the analysis process.
Suppose that each code segment $b$ has an activity classification $A_b$ of model training or data cleaning.
Suppose two data scientists are given the same overall task (e.g., train a model to predict flight delays). 
Methodological variation describes differences in the sequences of activity classifications for each data scientist.

\begin{figure}[ht!]
    \centering
    \includegraphics[width=0.5\linewidth]{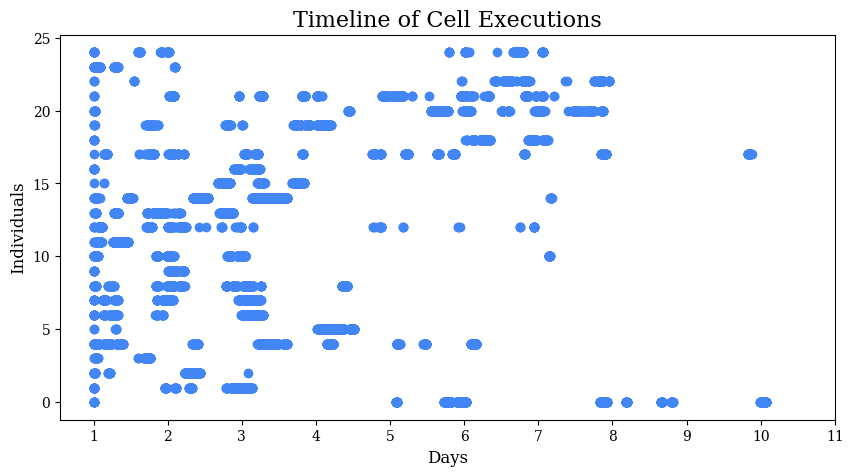}
    \caption{A timeline of cell executions in Flight v1}
    \label{fig:timeline}
\end{figure}

We include some example data from our Flight V1 dataset.
In Figure \ref{fig:timeline}, we can observe the timeline of when each student-executed cell runs relative to their first execution. The students generally worked on the project across multiple days, with significant breaks between sessions. There is a high variance between the time spent by each student. One could use data like this to characterize and correlate that variance with outcomes. Understanding methodological variation is crucial for systems researchers. Data science software systems are often optimized for particular workload patterns.

\subsection{Implementation}
\begin{figure}[ht!]
    \centering
    \includegraphics[width=0.6\linewidth]{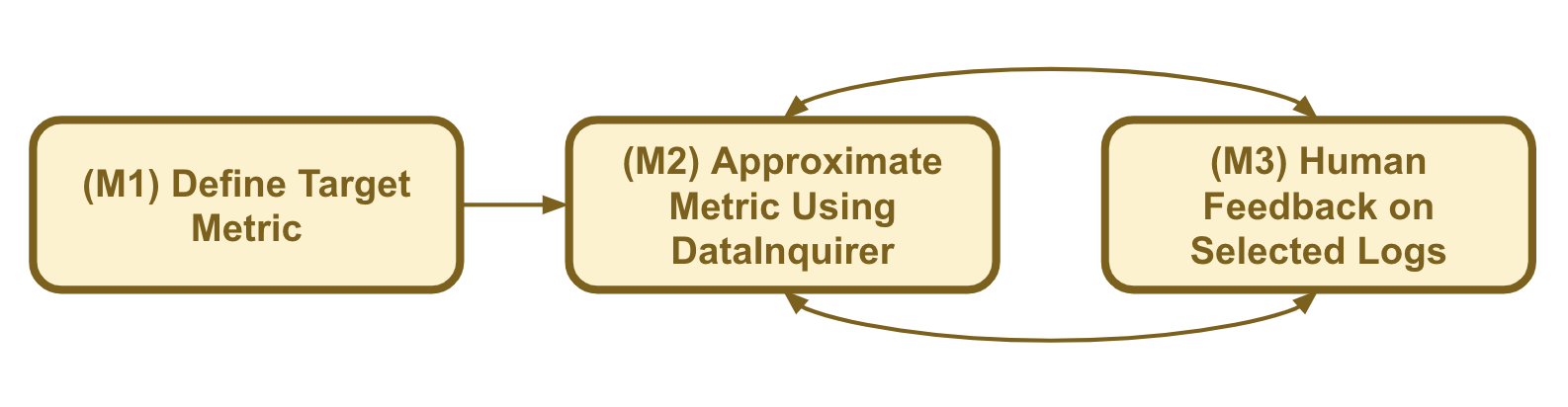}
    \caption{Procedure for analysis of variation over logs}
    \label{fig:method}
\end{figure}

In this section, we describe the procedure used to analyze variation in this work. We approached the problem using a human-in-the-loop procedure on top of \sys. Our evaluation process, as depicted in Figure \ref{fig:method}, follows a well-defined structure. All individual data analyses begin with a common first step: identifying the target metric for measurement from the available logs (M1). In our variation analysis, this metric could be the number of external libraries used, lines of code for visualization, or code lineage distance. The authors carefully chose these metrics after thoughtful discussion and deliberation. 

We iterate the procedure to capture this metric with human feedback. A hypothesis about this procedure is first created by the authors (M2). These are applied to the data logs using \sys. A subset of data logs is then selected, and we evaluate the result of the procedure on those logs (M3). If the results do not align with the human evaluation of the original metric, the measures are revised (M2). This process is repeated until the selected procedure sufficiently captures the essence of the original metric. We found that this process successfully captured the original metrics within our current log distribution. It was not designed to guarantee the applicability of these indirect measures to a wider range of logs.


\section{Results}
\subsection{Summary of Findings}
We organize the data analysis of the pilot study into six research questions.

\begin{itemize}
    \item \textbf{Research Question 1. (RQ1) Do different participants approach data cleaning in different ways (Semantic Variation)?} When given the same data science task on dirty data, participants identify different error types and use different methods to fix those errors.
    \item \textbf{Research Question 2. (RQ2) Do differences in data cleaning meaningfully change the conclusions (Semantic Variation)?} Using different data-cleaning methodologies is associated with divergent conclusions on at least one of the pilot study datasets.
    \item \textbf{Research Question 3. (RQ3) In terms of code structure and notebook usage, how do novices compare to experts (Syntactic Variation)?} We find that experts are more concise and faster in their approach to a data science task. In Flight v2, we find that the gap between novices and experts significantly narrows and postulates that this is associated with using LLMs for coding.
    \item \textbf{Research Question 4. (RQ4) How significant is the use of large language models in the Flight v2 dataset (Syntactic and Semantic Variation)?} We find that at least a third of participants chose to use LLMs when given the option.
    \item \textbf{Research Question5. (RQ5) In terms of effort spent on data science subtasks, how do novices and experts compare (Syntactic and Semantic Variation)?} Novices spend more effort on data modification than experts, and we notice this gap narrowing in the Flight v2 dataset.
     \item \textbf{Research Question 6. (RQ6) In terms of the time sequence of operations, how do experts and novices compare (Methodological Variation)?} Expert workflows tend to be simpler and more linear than those of novices.
\end{itemize}

\subsection{Variation Analysis of the Traffic Dataset}
We will start by presenting selected results on the Chicago Traffic Dataset. These results reflect a prototype implementation of \sys and are less complete than the subsequent results. Nonetheless, they are informative and motivate the problems that \ sys can solve.

\subsection*{RQ1. Do different students approach data cleaning in different ways?}
In our first analysis of the logs, we evaluate how different students approach data cleaning. The questions we asked the students encouraged them to think about semantic errors and missing values in the dataset. Despite receiving the same prompts, we found that the cleaning approaches differed. Below, we capture some key performance indicators of the students' data cleaning code. 

\begin{table}[ht!]
    \centering
    \begin{tabular}{@{}lccc@{}}
        \toprule
        & \textbf{Min} & \textbf{Median} & \textbf{Max} \\
        \midrule
        \textbf{Line of Cleaning Code} & 3 & 11 & 51 \\
        \textbf{\# Columns Referenced in Code} & 1 & 4 & 12 \\
        \textbf{\# Columns Referenced in Survey Results} & 4 & 6 & 16 \\
        \bottomrule
    \end{tabular}
    \caption{Summary statistics of students code}
    \label{tab:summary}
\end{table}

Table \ref{tab:summary} records the lines of code used for data cleaning, the number of distinct columns referenced in that code, and the number of distinct columns referenced in the student's explanation of the errors in the dataset. Students vary greatly in the complexity and scope of their data-cleaning operations. Some students have relatively short data-cleaning routines, while others have longer ones. Some clean only a few columns of data, while others clean the whole dataset. 

More interestingly, we find differences between the survey results and code measurements. Students often actually cleaned fewer errors than they reported in the survey. We hypothesize this is due to the code complexity or difficulty of fixing an observed error. These results underscore the importance of detailed code telemetry in data science user studies. Interview studies may yield aspirational responses from data scientists that do not fully reflect what they might implement in code.

\subsection*{RQ2. Do differences in data cleaning meaningfully change the conclusions?}
After analysis, we asked the students an open-ended question: ``Which neighborhoods in the city seem to have a higher frequency of crashes?''. We use this question to evaluate how data-cleaning choices may correlate with conclusions.
We went through the student responses to that question. We coded their responses into a few Chicago neighborhood categories: \textbf{West}, \textbf{West and Downtown}, \textbf{Downtown}, \textbf{South and Downtown}, \textbf{Unknown/Other/Incomplete}.
We associated the conclusion with the type of missing value operation the student uses (filling in or dropping missing values).
Table \ref{tab:data_statistics} shows the results.

\begin{table}[ht!]
    \centering
    \begin{tabular}{@{}lccccc@{}}
        \toprule
        & \textbf{West} & \textbf{West and Dtn} & \textbf{Dtn} & \textbf{South and Dtn} & \textbf{Unknown/Other} \\
        \midrule
        \textbf{Dropped Data} & 1 & 1 & \textbf{22} & 3 & 5 \\
        \textbf{Filled Data} & 2 & \textbf{7} & 11 & 0 & 3 \\
        \textbf{All} & 3 & 8 & 33 & 3 & 8 \\
        \bottomrule
    \end{tabular}
    \caption{Response to final survey question by cleaning operation}
    \label{tab:data_statistics}
\end{table}
While not statistically significant given the relatively small sample size, we found that students who imputed/filled in missing data were more likely to respond with the \textbf{West} region.
In other words, there is evidence that different data-cleaning operations can lead to different conclusions.
These results motivated us to construct a more detailed study with the flight dataset.

\subsection{Variation Analysis of Flight Datasets}
Next, we present an in-depth analysis of the flight delay prediction datasets. We present these results comparatively so they can be evaluated cross-sectionally and longitudinally.
These results should be contrasted with the previous ones. Rather than giving each participant a constrained, narrow task such as cleaning or missing value removal, they have a broader task. This allows for greater flexibility by participants to make high-level decisions.

\subsection*{RQ3. Regarding code structure and notebook usage, do novices vary more than experts?}
\begin{figure}[ht]
  \centering
  \includegraphics[width=0.32\linewidth]{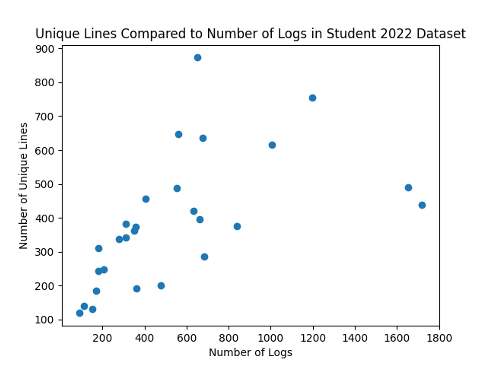}
  \includegraphics[width=0.32\linewidth]{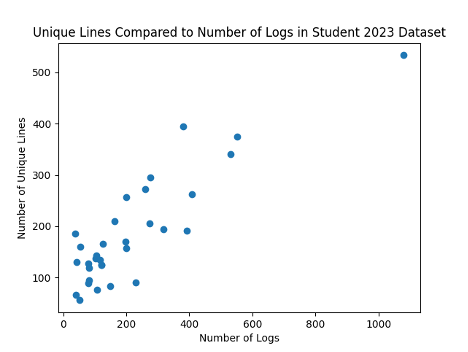}
  \includegraphics[width=0.32\linewidth]{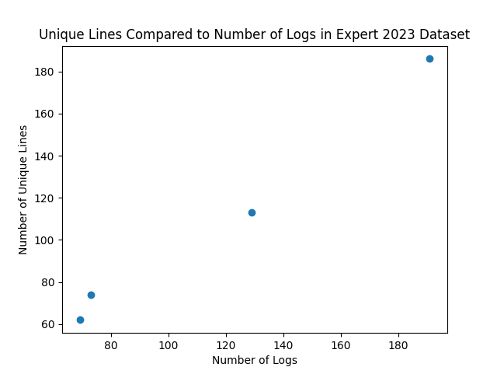}
  \caption{Experts generally write less code and execute it less}
  \label{fig:syntax_unique_lines}
\end{figure}

In this section, we focused on syntactical variation - indicators in logic processing independent of the code's content. Figure \ref{fig:syntax_unique_lines} shows the number of unique lines compared to the number of logs. Participants in the v3 dataset generated fewer logs, with fewer lines of code than the student datasets; likewise, participants in the v2 dataset generated fewer logs and fewer lines of code per log than those in the v1 dataset. This validates that there are differences in efficiencies between demographics in data science. We observe a linear correlation between the number of unique lines and the log count. This strong correlation suggests that a log increase indicates new work rather than simply repetition.

\begin{figure}[ht]
  \centering
  \includegraphics[width=0.3\linewidth]{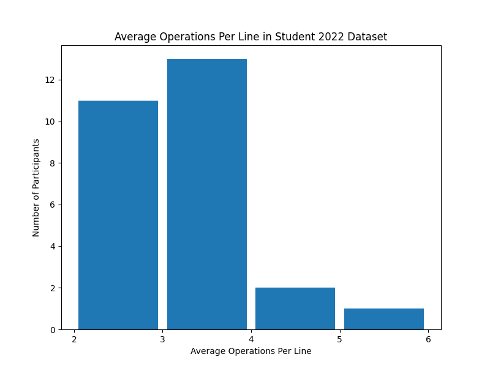}
  \includegraphics[width=0.3\linewidth]{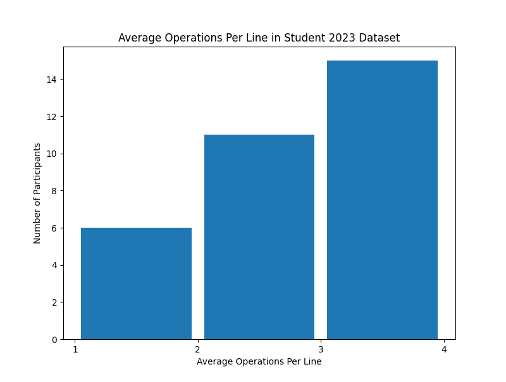}
  \includegraphics[width=0.3\linewidth]{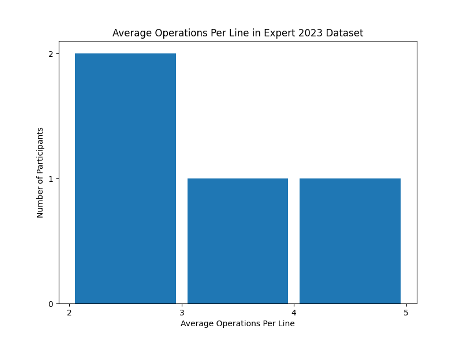}
  \caption{Not only do experts write less code, but their coding patterns seem to be of similar complexity to novices}
  \label{fig:syntax_average_op_expert}
\end{figure}

\begin{figure}[h]
  \centering
  \includegraphics[width=0.3\linewidth]{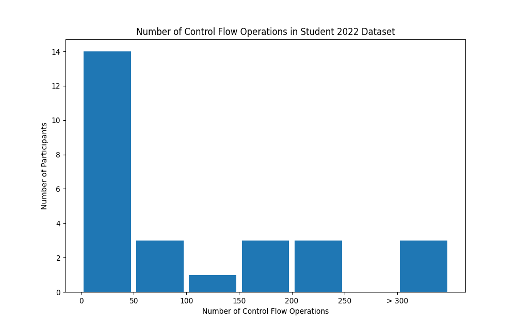}
  \includegraphics[width=0.3\linewidth]{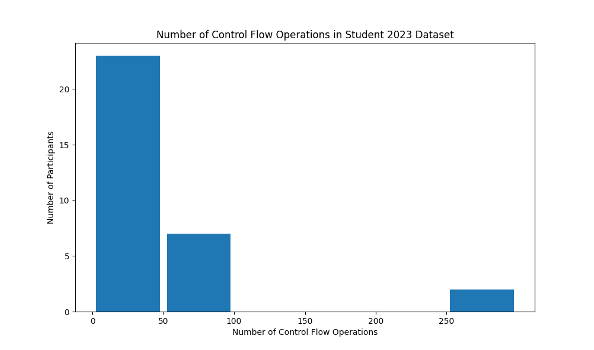}
  \includegraphics[width=0.3\linewidth]{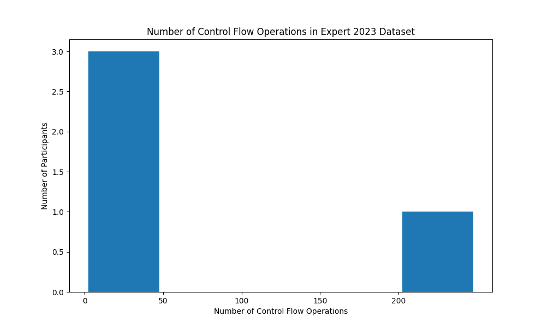}
  \caption{Most participants chose not to use control flow, but there are outlier participants who used them frequently}
  \label{fig:syntax_flow}
\end{figure}

One might attribute the previous result to code complexity, where novices write simpler code than experts. However, we show this is not true.
Figure \ref{fig:syntax_average_op_expert} shows the average number of operations per line across participants. We found that experts had roughly the same number of operations per line. We found that the average number of operations was consistent across all participants. Figure \ref{fig:syntax_flow} also shows the control flow variations between participants. We also approximated the operators through the search for Pythonic special words (\texttt{def},\texttt{if}, \texttt{for} etc.). Across all three datasets, there exist outlier participants that use significant control flow operations. This suggests that most participants chose not to use control, but the outliers that use control flow do so with high occurrences. Such users should be supported and accounted for in data science programming tools.

\subsection*{RQ4. How significant is the use of large language models in
the Flight v2 dataset?}
This section considers metrics explicitly related to large language use in the v2 dataset. For this set of participants, no restrictions were applied to the use of large language resources, but we asked the students to self-report their usage. The assignment was released in the Fall of 2023, which coincided with about one year after ChatGPT was released. While some specific metrics may have changed since then, this gives an interesting snapshot into LLM adoption in the academic setting. There are some key takeaways from our survey: (1) about a third of students self-reported that they did not use the tools, (2) most of the prompts given are related to code generation, and (3) there is significant variation in how students use LLM. 

\begin{figure}[h]
\centering
  \includegraphics[width=0.3\linewidth]{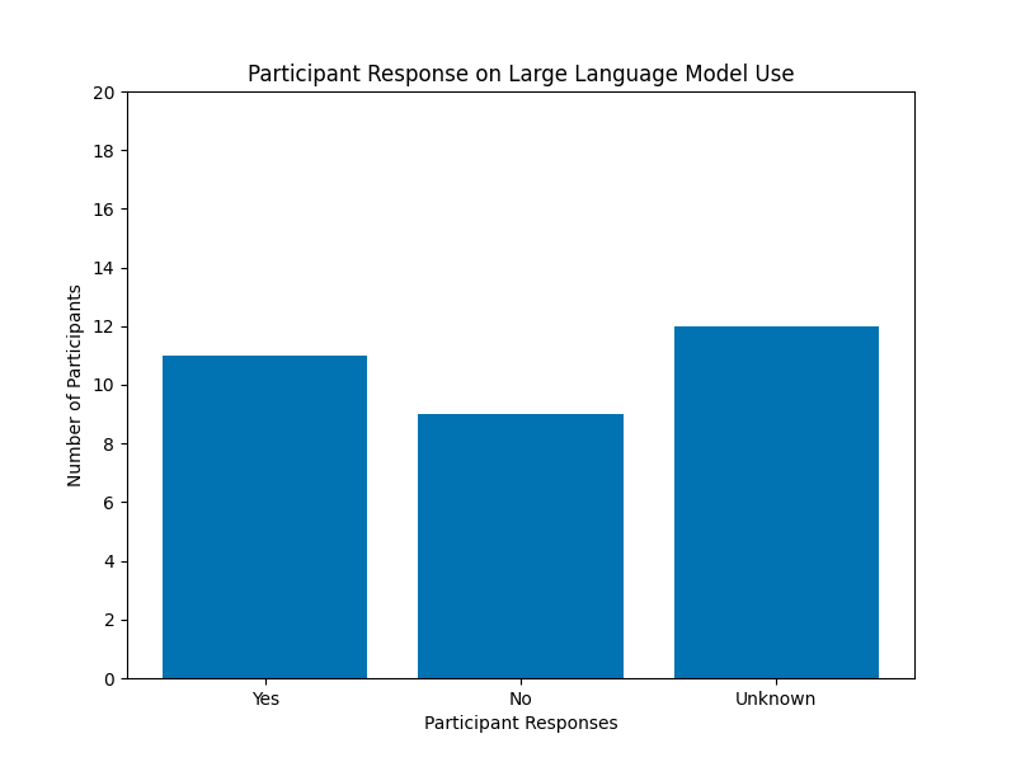}
    \caption{Self-reported use of large language models.}
  \label{fig:llm_response}
\end{figure}

We asked the students to self-report LLM usage in their Jupyter Notebook. This assignment task was not evaluated for completion; the data may be incomplete due to reliance on participants' self-reporting. The first question asked whether the participant used LLM, and Figure \ref{fig:llm_response} summarizes their response. The three response groups were split relatively evenly. We lack data to extrapolate usage for the group that did not respond and, therefore, cannot conclude the exact adoption percentage. However, at minimum, nine students stated that they did not use LLMs. Most of the discourse in academic teaching has been about adjusting to LLMs adoption, but there has been less discussion on how to support students who do not use these models and whether they are disadvantaged. We expected the response to this question to be binary. However, of the students that used LLMs, a couple presented justification for their usage (e.g., \textit{``Yes, but only to either debug set up or to explain functions''}). We believe this again reflects the discourse of LLMs in the academic setting; this study presents initial hints that suggest students self-regulate their usage - either from external pressures or from an internal desire to fulfill the core requirements of the assignment without assistance. 

Participants were requested to document the prompts used. We hand-labeled each prompt by their intended output: (1) Code Chunks. Generate multi-lined new code based on the requested purpose; (2) Code Modification. Given some initial code as a prompt, adjust this code for some purpose; (3) Line Code. Generate single-lined code; (4) Debugging. Fix a code error; (5)Explanation. Clarify or explain data science procedures.

\begin{figure}[h]
  \centering
  \includegraphics[width=0.3\linewidth]{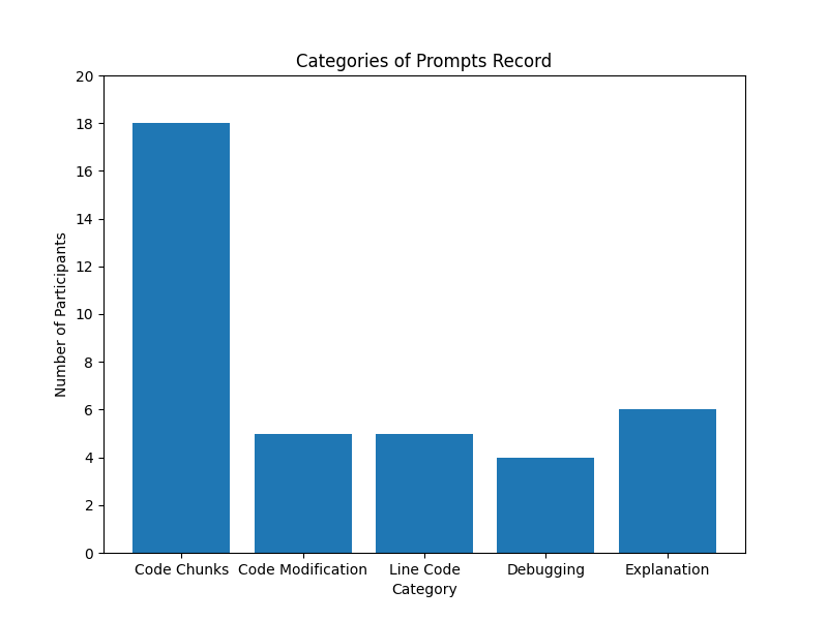}
  \caption{Category of prompts show that LLMs are largely used in end-to-end synthesis of code logic.}
  \label{fig:llm_category}
\end{figure}

Figure \ref{fig:llm_category} presents the distribution of these categories across all prompts. We noticed two key results - of all possible uses of LLMs for direct code interaction, code chunk generation significantly dominates the use cases. This suggests that LLMs are sufficient and largely used in the end-to-end synthesis of code logic from text. This has implementations concerning code literacy. In such cases, programmers may not practice understanding the API or syntax of code. Explanation-based prompts take up a relatively small part of the prompts recorded. 

Despite the largely syntactical nature of LLM questions, we found that some questions that students asked had a ``funneling'' effect where the LLM's results seem to greatly influence downstream outcomes. For example, we found prompts like the one below:
\begin{quote}
How to handle class imbalances in a machine learning training model in python 
\end{quote}
In 2023, we saw a number of students who employed a package \texttt{imblearn} that was not discussed in class. We found evidence linking the use of this package to LLM prompt results like the one above. Similarly, a number of students used the LLM to make data cleaning decisions:
\begin{quote}
I have a pandas dataset called "training" that has a specific column DelDep15. Can you write code to delete all NaN values from training dataset ONLY in DelDep15
\end{quote}
As we saw in RQ2, such questions can have significant impacts on downstream conclusions.
Using an LLM in the early stages of a data science project can funnel students down certain analysis workflows.
This gives an alternative explanation for the reduction in variation between Flight v1 and Flight v2 in RQ3 -- the use of LLMs reduces coding differences among participants. We also have reason to believe that self-reporting LLM usage was a likely under-reporting of overall student usage of LLMs.

\subsection*{RQ5. In terms of effort spent on data science subtasks, how do novices and experts compare?}

\begin{figure}[h]
  \centering
  \includegraphics[width=0.3\linewidth]{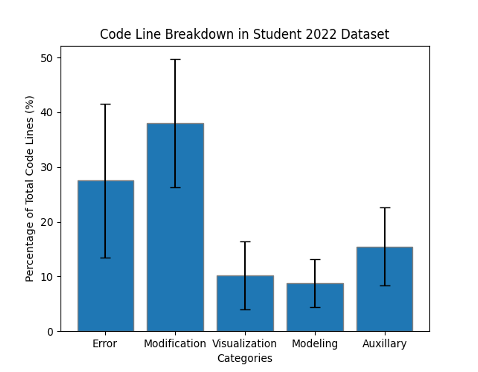}
  \includegraphics[width=0.3\linewidth]{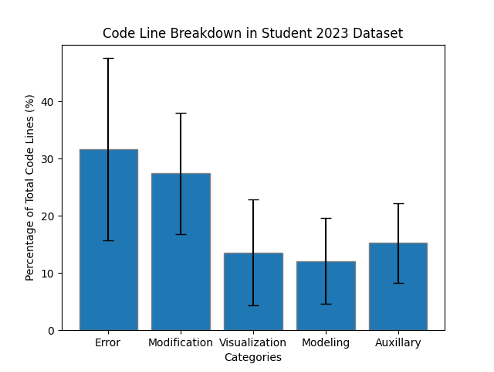}
  \includegraphics[width=0.3\linewidth]{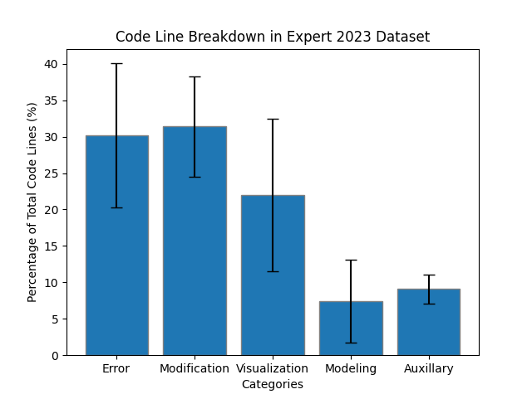}
  \caption{Distribution code categories show code for data modification make up the most significant}
  \label{fig:general}
\end{figure}

After manual observation of sample datasets by the paper's authors, we found that the participants' work can be generally categorized into standard data science categories: auxiliary code, data visualization, data modification, and modeling \cite{ramasamy2023workflow}. This follows from the problem presented to the participants. We additionally categorized code that generated output errors as a separate area for analysis. In this section, we independently analyze work within each category. Since those categories were not explicitly defined within the notebooks, a systematic effort was made to divide the student code. In our public dataset, each line of code includes a label tag for its category.

Figure \ref{fig:general} shows the distribution of these categories across datasets. The error bar shows the standard deviation between participants across datasets. Some interesting observations can be made from this figure. Firstly, this validates the claim that data cleaning makes the majority of work in data science - the combination of data modification and data visualization make up the bulk of code in the logs. In contrast, modeling lines compose a relatively small percentage. However, across the three datasets, error lines are either the largest or second largest category. This suggests that code debugging may take up a large proportion of the data process \cite{zhao_2023_hilda}. Visualization trends differ between the three datasets; experts generate noticeably more relative code for visualization. Similar to previous trends, the v2 dataset was closer to the expert distribution and had a higher percentage of visualization code on average. Overall, this categorization provides a quantitive view of the general effort spent in data science. This is an important statistic, and our evaluation match commonly stated results from user studies; i.e., most work in data science is spent in data cleaning \cite{anaconda2023state}. 

This breakdown creates a high-level overview of code semantics; we perform a more in-depth analysis of each category in the following sections. 
A consistent theme emerges from these results. While most of the statistical literature focuses on variation in modeling (e.g., model selection, feature selection, hyperparameter optimization), we consistently find that significantly more variation in upstream steps such as data modification and data exploration.
In other words, \textbf{data scientists vary more in how they scope a data science problem than in how they complete a data science problem}. 

\subsubsection{Auxillary Code}
\begin{figure}[h]
  \centering
  \includegraphics[width=0.3\linewidth]{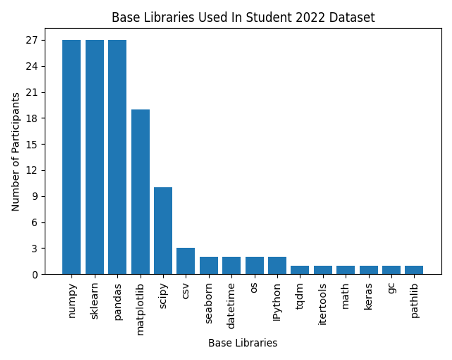}
  \includegraphics[width=0.3\linewidth]{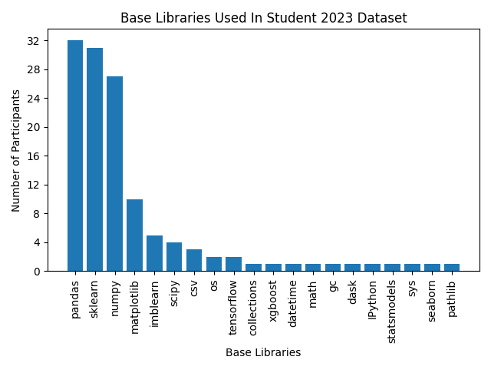}
  \includegraphics[width=0.3\linewidth]{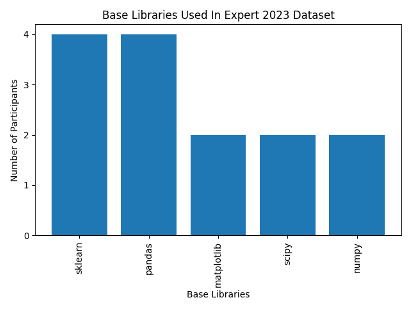}
  \caption{Usage of external libraries in participant workflows}
  \label{fig:import_blib}
\end{figure}

Figure \ref{fig:import_blib} shows the base import modules used by participants for all three datasets. The most common libraries are standard across all three datasets. The experts imported less base libraries. On the contrary, the v2 dataset had a longer tail of libraries. We suspect that large language models may be more efficient at recalling less frequently used libraries, leading to greater variation of libraries in data science. However, this can lead to a funneling effect where the LLM makes crucial design decisions for the data scientist.  

\subsubsection{Data Exploration}

\begin{figure}[h]
  \centering
   \includegraphics[width=0.3\linewidth]{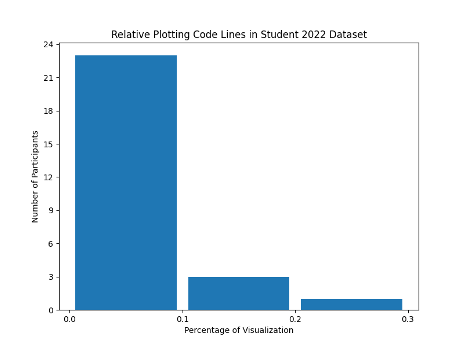}
  \includegraphics[width=0.3\linewidth]{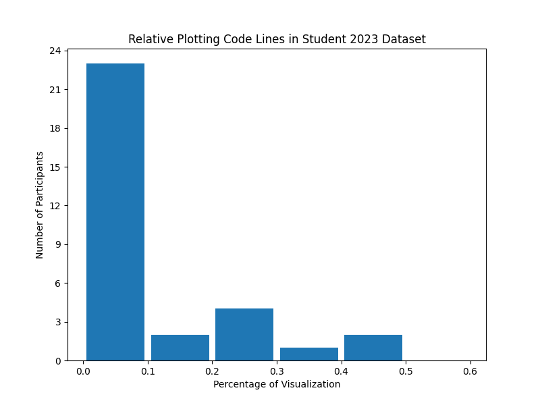}
  \includegraphics[width=0.3\linewidth]{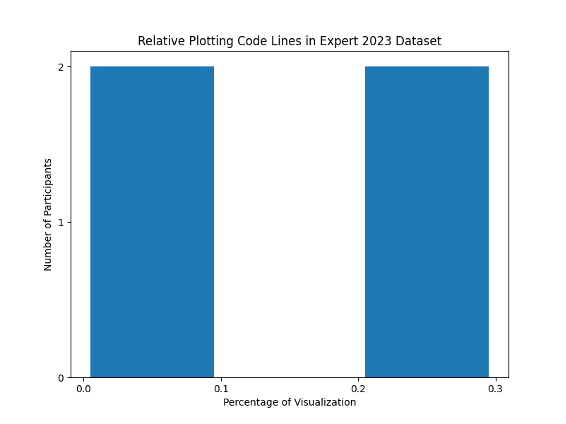}
  \caption{Plotting is a relatively small percentage of visualization}
  \label{fig:vis_cat}
\end{figure}

Figure \ref{fig:vis_cat} shows the relative percentage of plotting visualization by the participant out of the total visualization lines. Plots comprise a relatively small percentage of the total across all three datasets. In other words, direct ``print'' outputs are the bulk of visual outputs observed by the participants.

\subsubsection{Modeling}
\begin{figure}[h]
  \centering
  \includegraphics[width=0.3\linewidth]{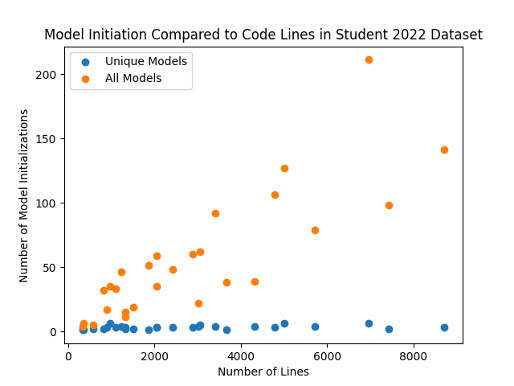}
  \includegraphics[width=0.3\linewidth]{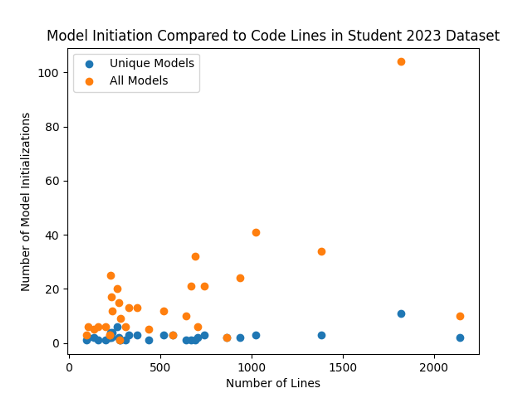}
  \includegraphics[width=0.3\linewidth]{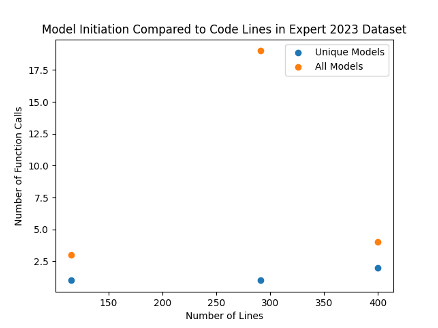}
  \caption{Number of unique models remain constant compared to total lines of code}
  \label{fig:model_lines}
\end{figure}

We focus on models in the \texttt{scikit-learn} library. While a couple of different libraries were explored, all final generated models used this library. Figure \ref{fig:model_lines} shows the number of model initiation lines compared to the total number of code lines. We observe that the pattern holds between datasets - the number of total model initializations increases with lines of code, but the number of unique models used is consistent. The number of unique models is also relatively small - most participants use less than ten models for this task. The type of models used by the participants were standard for tabular machine learning tasks - the most common ones being decision tree, random forest, logistic regression, and MLP classifier \cite{xin_paper_study}.

\begin{figure}[h]
  \centering
  \includegraphics[width=0.3\linewidth]{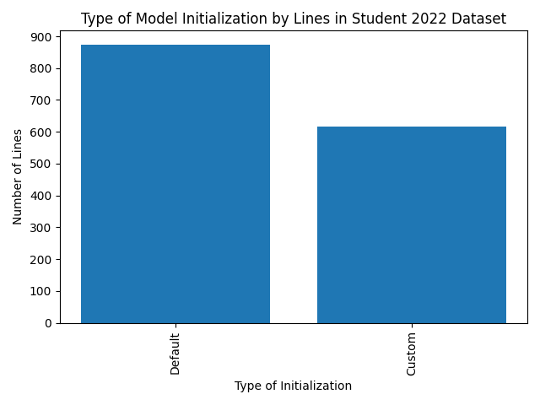}
  \includegraphics[width=0.3\linewidth]{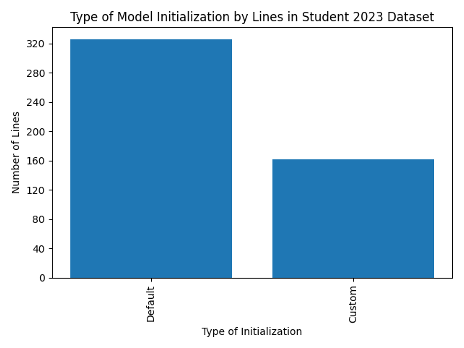}
  \includegraphics[width=0.3\linewidth]{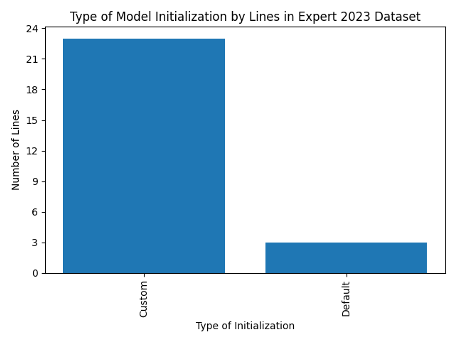}
  \caption{Data science experts are more likely to initialize models with custom parameters}
  \label{fig:model_type}
\end{figure}

In the \texttt{scikit-learn} library, all machine learning models are configured with adjustable parameters. For example, the RandomForestClassifier has parameters such as n\_estimators, which determines the number of trees in the forest, and max\_depth, which specifies the maximum depth of each tree. We split the type of model initialization based on whether the model uses custom or default. Figure \ref{fig:model_type} shows the distribution of model type by initiations. We observe that most expert initiations are custom, whereas student initiations tend to use default parameters. We theorize that experts would have more experience with the models and be more opinionated on the particular settings. This demonstrates variations between the demographics of data science participants.

\subsubsection{Errors}

\begin{figure}[h]
  \centering
  \includegraphics[width=0.3\linewidth]{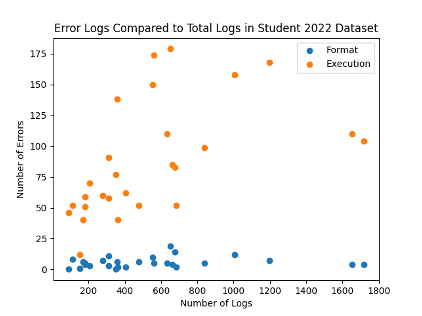}
  \includegraphics[width=0.3\linewidth]{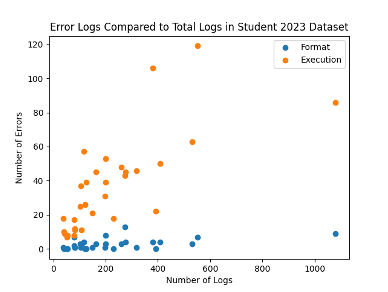}
  \includegraphics[width=0.3\linewidth]{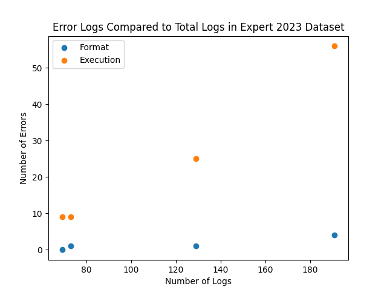}
  \caption{Execution errors increase with number of logs, but format errors stay constant}
  \label{fig:error_lines}
\end{figure}

\ref{fig:error_lines} shows the number of error logs compared to the total number of format and execution errors. Format errors in Python are errors caused by incorrect syntax, whereas execution errors are errors that occur when the program is running. We observe little variation between the participants regarding the number of format errors. Surprisingly, format errors are not correlated with the number of logs and are a small fraction of total errors. However, there is a positive correlation between execution errors and the number of logs. We suspect this could be causal in both directions - more experimentation might lead to more execution errors, and more execution errors may require more log iterations to correct.

\begin{figure}[h]
  \centering
  \includegraphics[width=0.3\linewidth]{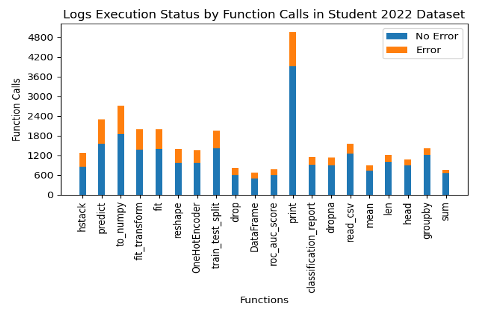}
  \includegraphics[width=0.3\linewidth]{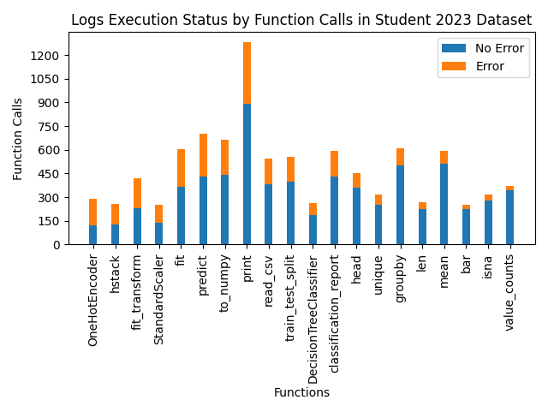}
  \includegraphics[width=0.3\linewidth]{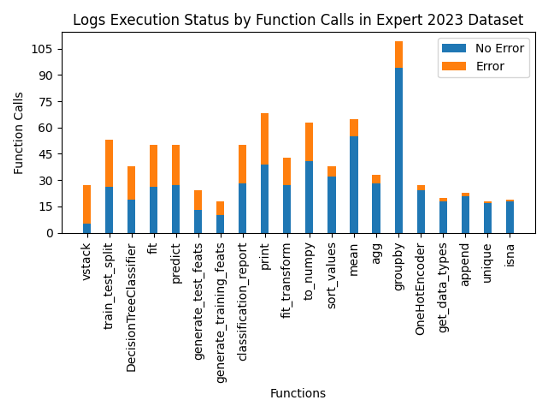}
  \caption{Some functions may have high correlation with execution errors}
  \label{fig:error_funct}
\end{figure}

We additionally investigated the correlations between particular function statements and errors. For each dataset's top 20 most frequently called functions, we graphed the number of function calls that resulted in errors or successful executions. Figure \ref{fig:error_funct} shows the results, sorted by the decreasing ratio of errors. We observe patterns in this data across all three datasets; functions related to the model training (e.g., \texttt{fit, predict, and fit\_transform}) tend to have higher error ratios, occurring in the top ten functions. On the other hand, simple data manipulation functions (e.g., \texttt{groupby, mean, head, isna}) showed lower error ratios, all occurring in the bottom ten functions. Understanding error patterns and variations can allow more impactful ways to reduce the burden of debugging across data science projects.

\subsubsection{Data Modification}

\begin{figure}[h]
  \centering
  \includegraphics[width=0.3\linewidth]{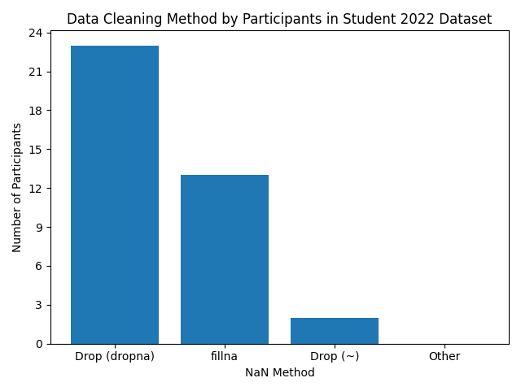}
  \includegraphics[width=0.3\linewidth]{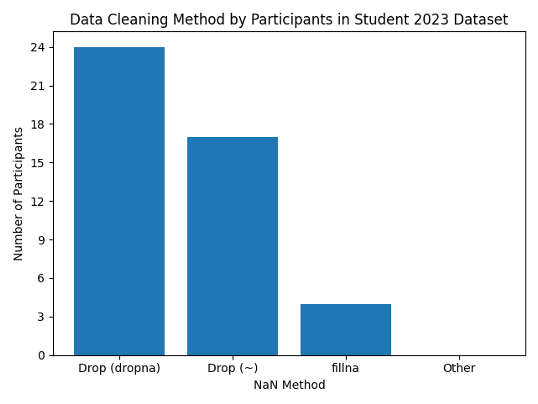}
  \includegraphics[width=0.3\linewidth]{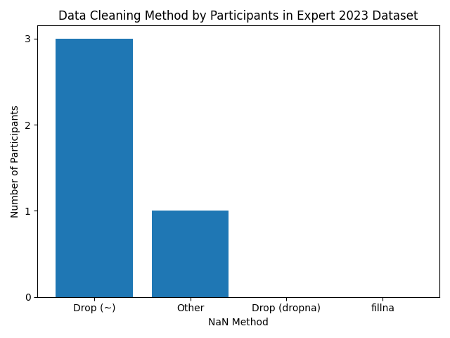}
  \caption{Most participants simply dropped the samples with missing data}
  \label{fig:mode_clean_method}
\end{figure}

We specifically focused on how participants dealt with missing data, similar to our research question over the chicago dataset (\textbf{RQ2}). We found again that the majority of participants followed two conventions - either the samples with missing data were dropped, or the missing data was filled with a constant value. When missing data is dropped, two fixed techniques were used: calling the \texttt{dropna} function from the \texttt{pandas} library or directly removing samples with missing data with the ~ operation. In Figure 
\ref{fig:mode_clean_method}, we show the distribution of missing techniques by participants across all datasets. This figure shows that simply dropping the data samples is the most common approach; there is low variation in techniques for this data cleaning task. 

This is not the best procedure for dealing with non-random missing values. In the original data, the missing data is not random but indicates canceled flights. We examined the v3 dataset and found only one expert found this pattern - they explicitly commented: \textit{``most of the nan values are cancelled in some way!''}. This is an interesting result since it suggests that the simplest data cleaning was adopted for our Flight datasets.

\subsubsection{Summary}
Overall, we outline a diverse set of findings for \textbf{RQ5} that give insight into semantic variation. We believe these findings may have great impact on future practices related to data science. Capturing detailed practical patterns and variation in how data scientists perform tasks within an end-to-end workflow can lead to better system support for these workflows. For example, knowing that LLMs may increase the diversity of external libraries in data science allows practitioners to justify more effort on the robustness of these libraries. Similarly, understanding how error patterns correlate with function calls can lead to more intuitive implementation of those functions. 

Additionally, capturing how data science is implemented in practice can help address gaps between novices and best practices. In future courses, we can focus on how exploration can help discover non-random patterns in missing data. We can also given more focused practical assignments for model-tuning that allow students to gain intuition for how hyper-parameters affect model behavior.

\subsection*{RQ6. In terms of methodology (or the time sequence of operations), how do experts and novices compare?}

\begin{figure}[h]
  \centering
  \includegraphics[width=0.3\linewidth]{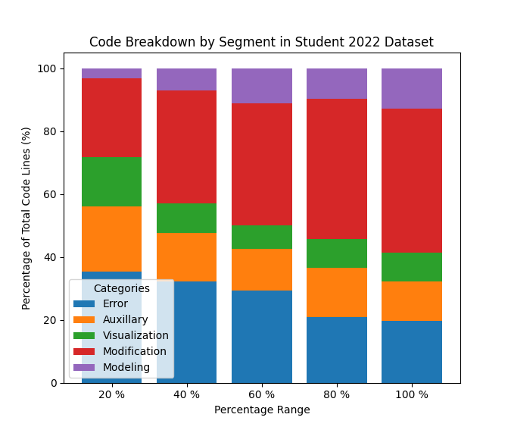}
  \includegraphics[width=0.3\linewidth]{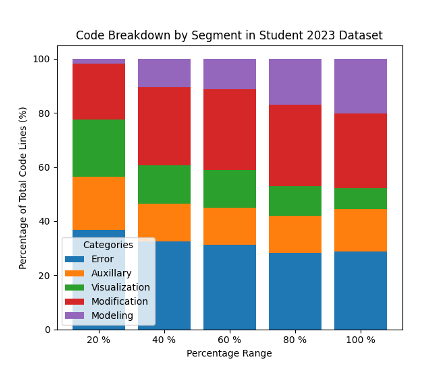}
  \includegraphics[width=0.3\linewidth]{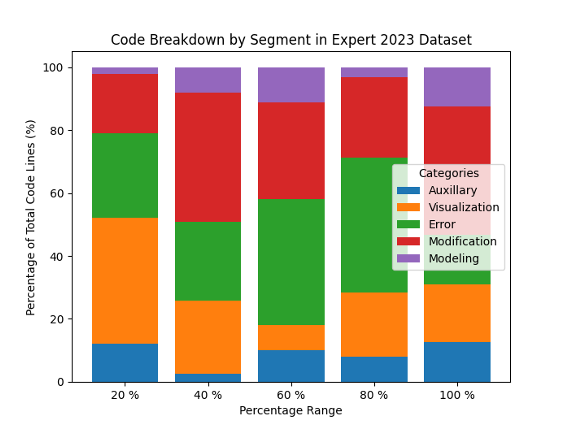}
  \caption{Distribution of code categorization in different segments of the log}
  \label{fig:process_segment}
\end{figure}

To evaluate the code progression of the data, we divided each log history into ten even segments and evaluated the distribution of code categorization within each segment. Figure \ref{fig:process_segment} shows this distribution averaged across participants for each dataset. We observe that there are not many clear trends in the distributions. One such trend is that model lines are more frequent as the log progresses, for obvious reasons. This suggests that the category definition was relatively broad, and a mixture of all categories is present throughout the process. 

\begin{figure}[h]
  \centering
  \includegraphics[width=0.3\linewidth]{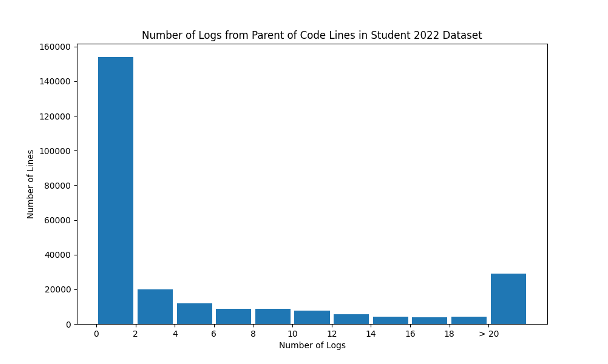}
  \includegraphics[width=0.3\linewidth]{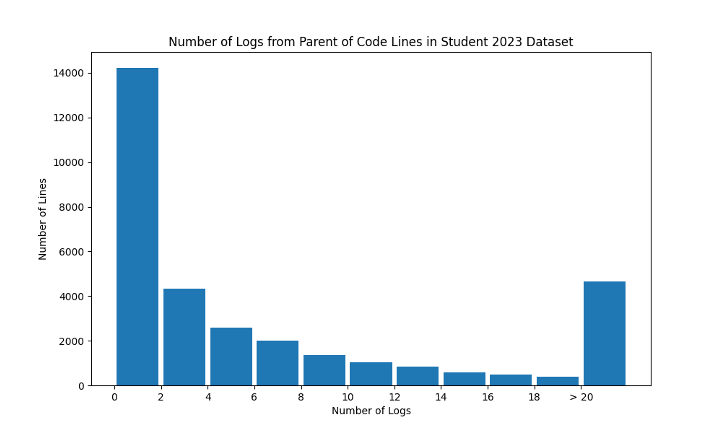}
  \includegraphics[width=0.3\linewidth]{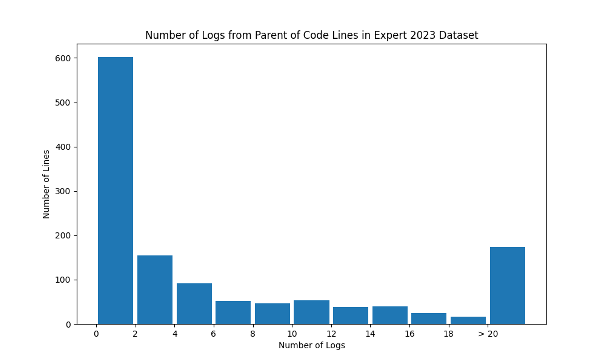}
  \caption{The lineage of the code lines is mostly linear}
  \label{fig:process_lines}
\end{figure}

\begin{figure}[h]
  \centering
  \includegraphics[width=0.3\linewidth]{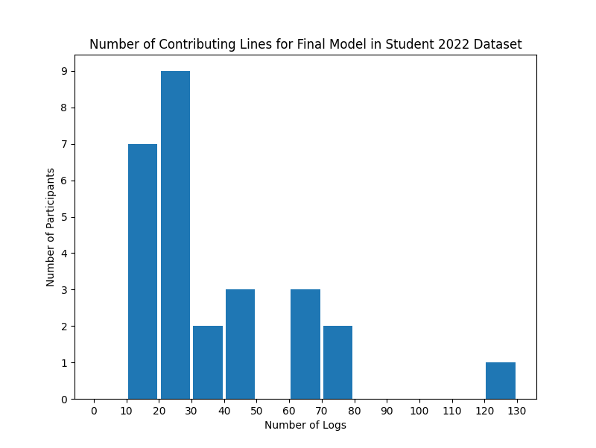}
  \includegraphics[width=0.3\linewidth]{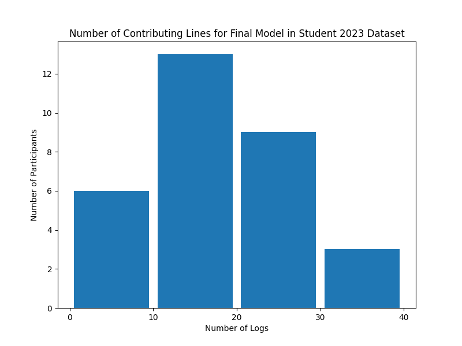}
  \includegraphics[width=0.3\linewidth]{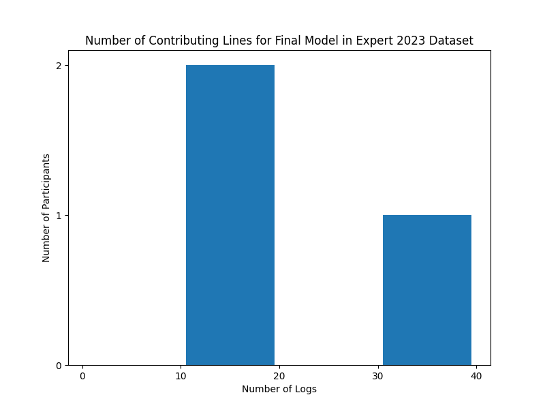}
  \caption{There are relatively few direct contributing lines for the final model}
  \label{fig:process_contribute}
\end{figure}

Next, we evaluated metrics related to the lineage of the dataset. For each code line, Figure \ref{fig:process_contribute} shows the distribution of the distance between parent children lines. In most cases, the parent line is within the same log or from the previous log. However, similar to previous patterns of variance, there is a long tail of outliers. Additionally, we evaluated the lineage of the final model. Figure \ref{fig:process_lines} shows the number of direct contributing lines that final model depends upon. Firstly, we observe differences between the datasets - the v1 dataset has a higher number of and higher variance of lineage lines. Overall, direct contributing lines for this model are relatively few compared to the total number of lines, suggesting that the final generated pipeline is short and that the bunk of the work is exploration.

\subsubsection{Handling Errors}
\begin{figure}[h]
  \centering
  \includegraphics[width=0.3\linewidth]{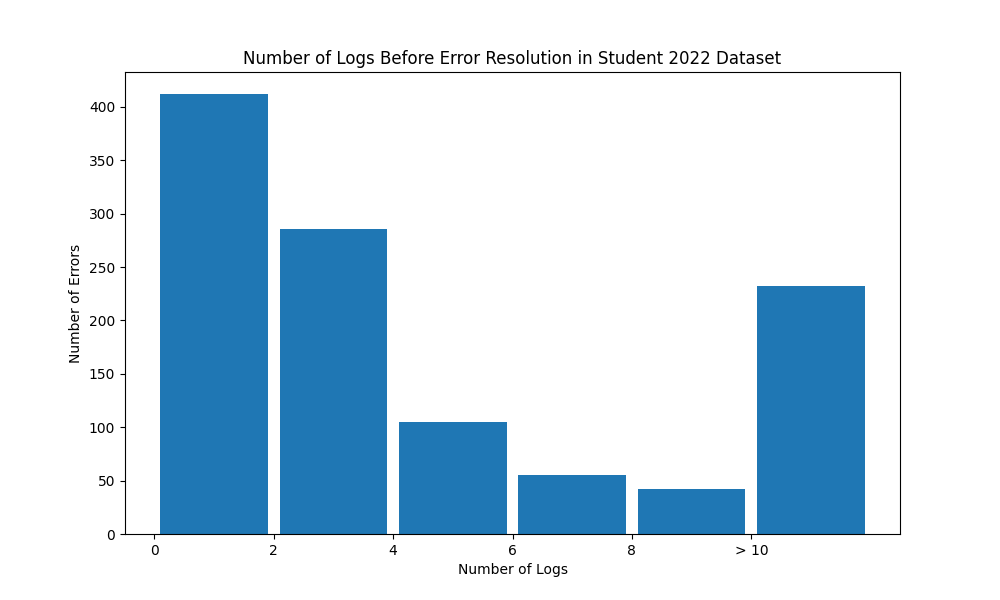}
  \includegraphics[width=0.3\linewidth]{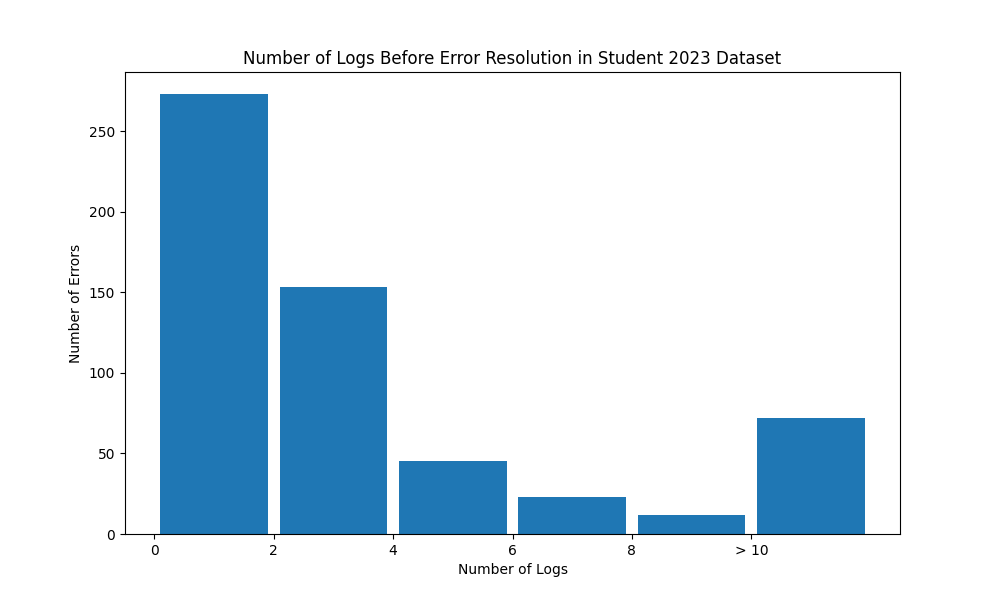}
  \includegraphics[width=0.3\linewidth]{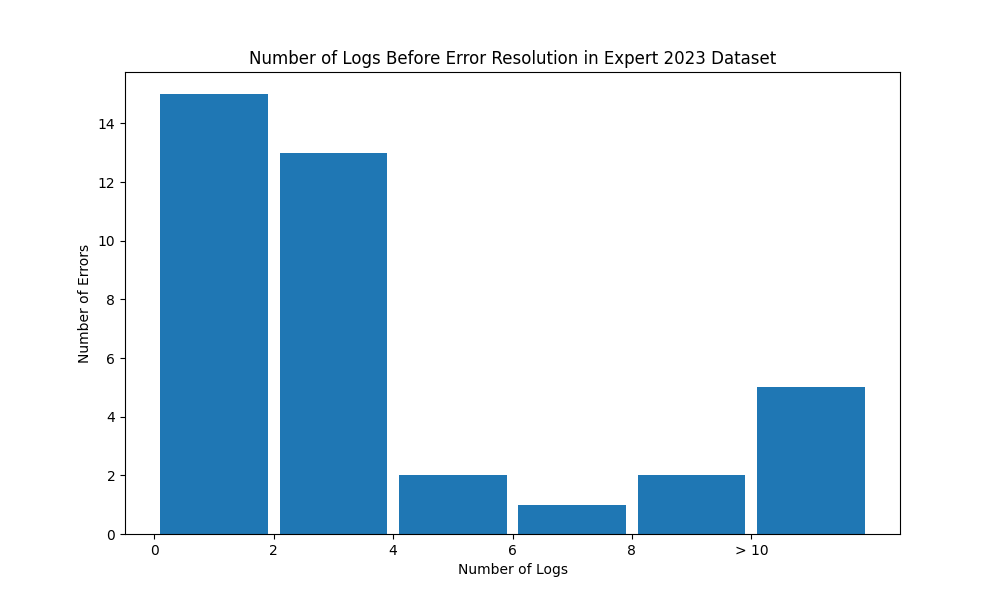}
  \caption{Most errors are resolved within four logs}
  \label{fig:process_elog}
\end{figure}

Now, let us explore the process of error resolution. We evaluate error resolution by using the code's vector similarity. Given an initial occurrence of an erroneous log, we deem that an error is resolved if a similar log later in the log history does not generate errors. We update the original code with the last log that generates errors to prevent code drift. This process allows us to analyze the debugging process. We found that the majority of errors are resolved. Figure \ref{fig:process_elog} shows the number of logs before error resolution. We observe expected behavior - errors tend to be resolved within four logs of the original error, but there is a tail of high logs. The tail from the v2 dataset is smaller than the tail from the v1 dataset. We hypothesize that although this was not fully reflected in the student responses, large language models help reduce the length of the debugging process.

\section{Discussion}

\subsection{Related Work}

Interview studies have been made to profile data science workflows \cite{kery_story, shankar_2022, kandel_2012, whither, muller2019, Liu2019PathsEP, Kross2019PractitionersTD, anaconda2023state}. These studies have limitations due to human bias when self-reporting their work \cite{bias}. Muller et al. captured and classified human interactions with data through various user studies with data workers at IBM \cite{muller2019}, specifically exploring data discovery, capture, curation, design, and creation. They found that data science is a \textit{craft} and depended on human expertise to capture patterns. Similarly, Liu et al. found that data scientists take in multiple considerations when making decisions and have varied judgment calls concerning particular methodologies and paths taken \cite{Liu2019PathsEP}. Kross et al. explore data science education and outline how current teachers consider diverse backgrounds, technical challenges, and data uncertainty in how they structure their teachings\cite{Kross2019PractitionersTD}. For the last six years, Anaconda, a Python software company, has released yearly large-scale industry surveys on data science that capture current trends, including in external library adoption, generative machine learning, and business value\cite{anaconda2023state}.

Works have attempted to quantify data science behavior, often through large-scale analyses of notebooks in online repositories like GitHub \cite{Rehman2019TowardsUD, Rule2018ExplorationAE, Pimentel2019ALS, microsoft_2019}. However, these approaches are limited by the unknown context of the notebooks, such as the intended purpose and authorship. Additionally, prior research has largely overlooked variation among data scientists. Our work addresses these gaps, offering unique analyses of individual variation and detailed task-based differences between participants. Rule et al. explored notebooks on GitHub, focusing specifically on those from academic settings, to explore the tension between exploration and explanation \cite{Rule2018ExplorationAE}. Pimentel et al. studied the reproducibility of Github notebooks and concluded that only about 20\% of notebooks are reproducible due to various errors \cite{Pimentel2019ALS}. A large-scale study by Psallidas et al. showed some similar results to our paper - there is a long tail of library imports, and data science workflows are largely linear \cite{microsoft_2019}. 

To the best of our knowledge, there has yet to be a detailed study that quantitatively captures fine-grained code analysis over the full history of user behavior. Lee et al. explore iterations in machine learning model training on OpenML workflows but only consider the model training step rather than the data science process as a whole \cite{lee_2020}. Xin et al. present early work quantifying machine learning iterations based on secondary research paper sources \cite{xin_paper_study}. Raghunandan et al. studied the progression of explanation vs. exploration in notebooks, using a regression model on general notebook characteristics (i.e., number of markdown vs. code cells). Ramasamy et al. explored general semantic and methodology behavior, taking a machine learning approach to categorize notebook cells, and showed that data science steps may be iterative\cite{Ramasamy2022WorkflowAO}. Pre-large language models, Koenzen et al. studied duplication in notebooks, showing that self-duplication occurs in small numbers, but duplication generally occurred from online resources\cite{Koenzen2020CodeDA}.

There has been a significant discussion on the topic of large language model usage in the classroom \cite{Prasad2023GeneratingPT, Zheng_2023, Macneil2022TheIO, Gan2023LargeLM, Meyer2023ChatGPTAL, Yan2023PracticalAE}, and for computer science \cite{Kalliamvakou_2024, Ziegler2024MeasuringGC, Zhou2023ExploringTP, Zhang2023PracticesAC, Tian2023IsCT, Jalil2023ChatGPTAS}. Most studies centered on how educators and practitioners should respond to the adoption of these models, with few works on the current usages in data science education in practice. For general coding tasks, Github’s user study showed that users completed programming tasks 55\% faster with CoPilot \cite{Kalliamvakou_2024}. Prasad performed a user study on language model usage in an advanced formal method course in the spring of 2023 that showed students used the models sparingly, primarily for formal specifications \cite{Prasad2023GeneratingPT}. Zheng created specialized tasks for students involving ChatGPT prompts; his post-hoc user survey suggested that the tool is especially effective for coding in data science \cite{Zheng_2023}.

\subsection{Discussion of \sys}

\subsubsection{Lessons Learnt}

Over the three years where we conducted experiments in data science, the capture and analytics method for logs has greatly evolved. We highlight some key observations that were made during this process. Firstly, participant behavior using Python files and Jupyter notebooks greatly differed. This insight is consistent with previous discussions on notebook behavior \cite{Rule2018ExplorationAE}. Therefore, it was important to incorporate and consider the appropriate environment when conducting the study. We focused on supporting notebooks since that is the preferred environment for data science \cite{microsoft_2019}.

In the second iteration of \sys, we implemented a binary capture system that significantly enhanced the robustness and integrity of the logging process. By automatically ceasing logging upon detecting any failure and concealing the logging interface from participants, we minimized the risk of data tampering and reduced the incidence of unnoticed logging failures. These improvements have made \sys a more reliable tool in experimental settings, ensuring that data collection remains secure and consistent throughout the research process.

We found that the process of analyzing the data was iterative. After collecting the dataset, new questions naturally arose that required new types of analysis - our semantic query module was conceived and implemented after our pilot studies. This highlighted the importance of log replay to the governance and analysis of these data science workflows; replay introduces new opportunities to gather information not considered in the initial log. In addition, information that was not practical to store, such as cell output, could only be captured through replay. 

\subsubsection{Future Applications}

There has been a move toward adopting hosted cloud services for notebook \cite{colab}. This migration to the cloud opens up opportunities for a unified logging system where all activities and changes across notebooks and users are tracked in a single, secure location. If we extend these cloud services with the functionalities of \sys,  we can improve the monitoring, auditing, and debugging of workflows in data science. 

There are many possible applications for performing post-hoc analysis of workflows in data science and notebooks. However, the complexity of the log data presents a significant challenge to extracting meaningful insights. Adopting a standardized logging format and a centralized analysis framework will greatly expedite the process of future applications.

Our current implementations for the queries and replay libraries are simple. While we did not find performance to be an issue given the current volume of logs, as this volume increases, the response latency can hinder analysis \cite{Liu2014TheEO}. There is work to optimize our current implementations with more efficient data structures, parallel processing techniques, and caching mechanisms. This would ensure our system remains scalable and responsive as log volumes grow.

\section{Conclusion}
In conclusion, there are a wide range of benefits in an end-to-end system for execution logs of code files and computational notebooks. We introduce such a system, \sys, that provides a comprehensive platform for tracking and analyzing the incremental code executions of data scientists. \sys offers valuable insights into programming patterns, ultimately contributing to a deeper understanding of data science.

In a series of pilot studies gathered using \sys, 97 traces of data science activity were gathered using our system. These traces captured how participants approached precise data-related questions on the Chicago dataset, and end-to-end machine learning prediction on the Flight dataset. Analyzing these traces revealed that there is significant syntactic, semantic and procedural variation between different participants. By quantifying these diverse approaches, \sys paves the way for more informed and targeted improvements in data science education and tools.



\bibliographystyle{abbrv}
\bibliography{refs2}

%



\end{document}